\documentclass[letterpaper, 10pt]{article}
\usepackage[dvipsnames]{xcolor}
\usepackage{preprint}

\title{Neural Inference of Fluid--Structure Interactions from \\ Sparse Off-Body Measurements}

\author{
  Rui Tang$^1$,
  Ke Zhou$^1$,
  Jifu Tan$^{2}$, and
  Samuel J. Grauer$^{1,}$\thanks{Corresponding author: \href{mailto:sgrauer@psu.edu}{sgrauer@psu.edu}}\vspace*{.3em}\\
  {\small $^1$Department of Mechanical Engineering, Pennsylvania State University}\vspace*{-.25em}\\
  {\small $^2$Department of Mechanical Engineering, Binghamton University}\vspace{-1em}}
\date{}

\begin{document}

\maketitle
\setcounter{footnote}{2}
\vspace*{-2em}

% Abstract
\begin{abstract}
\noindent We report a novel physics-informed neural framework for reconstructing unsteady fluid--structure interactions (FSI) from sparse, single-phase observations of the flow. Our approach combines a modal surface model with coordinate neural representations of the fluid and solid states, constrained by the fluid's governing equations and interface conditions. Using only off-body Lagrangian particle tracks and a moving-wall boundary condition, the method infers both flow fields and structural motion. It does not require a constitutive model for the solid or measurements of surface position, although including these can improve performance. We demonstrate the approach numerically on two canonical FSI benchmarks: vortex-induced oscillations of a 2D flapping plate and pulse-wave propagation in a 3D flexible pipe. We also demonstrate it on flow around a swimming fish. In all cases, the framework achieves accurate reconstructions of flow states and structural deformations despite acute data sparsity near the moving interface. A key result is that reconstructions remain robust to over-parameterization. This work extends physics-informed neural networks to coupled fluid--structure dynamics learned from single-phase observations, and it provides a pathway toward quantitative FSI analysis when flow measurements are sparse and structural measurements are asynchronous or unavailable.\par\vspace*{.5em}

\end{abstract}
\vspace*{2em}

%%% Introduction %%%
\section{Introduction}
\label{sec: introduction}
Fluid--structure interactions (FSI) arise when a flexible structure is immersed in, or contains, a flowing fluid. They appear in a wide range of applications, including flapping-wing aerodynamics \cite{Nakata2012, Zhu2021}, offshore structures and wind turbines \cite{Calderer2018, Korobenko2013}, cardiovascular systems \cite{Suito2014}, and flow-induced vibrations \cite{Lee2017}. Accurately resolving these interactions is essential for understanding the underlying physics and improving the performance of engineering devices. Traditional computational methods for FSI, such as arbitrary Lagrangian--Eulerian (ALE) \cite{Donea1982, Hughes1981}, level set \cite{Legay2006, Jenkins2015}, fictitious domain \cite{Yu2013, Pathak2016}, and immersed boundary methods \cite{Peskin2002, Sotiropoulos2014}, have enabled high-fidelity simulations by capturing moving boundaries and coupling fluid and solid domains. Each approach offers trade-offs between numerical stability, geometric flexibility, and computational cost. These methods work well for systems with well-defined boundary conditions and known material models \cite{Hou2012, Haeri2012}, but their application to real-world FSI often faces serious limitations. In particular, inflow\slash outflow boundary conditions and material properties may be unknown, and resolving small-scale dynamics near a complex interface can be prohibitively expensive.\par

In such cases, experimental measurements are essential, but the data are often sparse, noisy, and limited to a single phase (typically the fluid). This creates a disconnect: simulations provide the rich spatio-temporally resolved fields needed to analyze FSI dynamics, yet they can be sensitive to uncertain boundary conditions and constitutive models. Experiments, on the other hand, reflect the actual system behavior but provide incomplete information. To leverage the strengths of both, we propose a data assimilation (DA) method that uses sparse, off-body measurements of the fluid to reconstruct both the flow field and the structural response, without requiring a constitutive model for the solid or direct structural measurements. Our approach combines the fluid governing equations with partial observations to produce physically consistent reconstructions in settings where conventional simulations are difficult to apply due to uncertain boundary conditions or material models. This capability is particularly relevant for FSI with complex interfaces or biological materials, where structural properties are poorly characterized and simultaneous two-phase measurements are often impractical.\par

There are two major strategies for simultaneously measuring the fluid and solid phases in FSI \cite{Chatellier2013, Bleischwitz2017}. The first combines two separate modalities, each with distinct illumination and\slash or detection wavelengths. For example, synchronous particle image velocimetry (PIV) and digital image correlation (DIC) have been used to study hydrodynamic loading and structural response \cite{Zhang2019}, jet interactions with compliant surfaces \cite{Hortensius2017}, and supersonic panel flutter \cite{DAguanno2023}. While these setups can provide high-resolution measurements of both phases with limited cross-talk, they are costly and require complex calibration. To reduce experimental overhead, a second strategy uses a single measurement modality, typically optimized for the fluid phase, and segments the fluid and solid in post-processing. For instance, PIV \cite{Kosters2023} and Lagrangian particle tracking (LPT) \cite{Mitrotta2022, Schroder2023} have been used to capture both flow and surface motion by embedding particles on the structure. These surface particles can be distinguished from fluid tracers based on spatial location (e.g., lying on the outermost surface) or velocity (e.g., moving more slowly than the advected particles). However, such single-modality approaches face challenges, including optical interference, complex data processing, and reduced effective resolution, since fewer particles are available per phase due to practical tracking limits. The present work develops an inversion algorithm tailored to this second class of experiments (while also being applicable to the first), using single-phase measurements to infer the flow and structural dynamics. The same strategy may also be extended to other non-invasive modalities such as magnetic resonance velocimetry (MRV) \cite{Elkins2007}.\par

Several algorithms have recently been developed to reconstruct aspects of FSI from single-phase data. For instance, Kontogiannis et al.~\cite{Kontogiannis2022} reported a Bayesian method for joint flow reconstruction and boundary segmentation from noisy MRV data. They modeled the 2D domain boundary using a signed distance function and employed an adjoint-based solver to optimize the inflow condition and boundary shape, subject to a hard constraint on the governing equations. Their approach yielded accurate reconstructions for steady 2D flows with fixed boundaries. Karnakov et al.~\cite{Karnakov2024} used optimizing a discrete loss (ODIL), a grid-based inverse solver that minimizes a composite loss function with soft constraints on the measurements and governing equations. ODIL minimizes the loss with a Newton solver and, in some settings, can recover flow and structure states from sparse velocity data while leveraging convergence properties of traditional discretized solvers. For FSI problems, the ODIL loss penalizes residuals from the discretized Navier--Stokes equations and a no-slip boundary condition, with additional regularization applied to the level set representation of the surface. However, reported results indicate that the method can struggle with non-convex or geometrically complex shapes, tending to approximate them by simpler convex forms, and its computational cost can grow rapidly with domain size and time in large-scale 3D or unsteady settings. More recently, Buhendwa et al.~\cite{Buhendwa2024} extended ODIL by integrating it into the JAX-Fluids solver to infer 3D obstacle shapes and flow fields in steady supersonic flows. By using explicit shock-capturing schemes and level set representations of the boundary, they demonstrated accurate shape reconstructions even in shock-dominated regimes. Together, these methods provide a useful foundation for inverse problems involving flows with uncertain boundaries. At the same time, unsteady problems with deforming interfaces and complex geometries remain challenging, particularly when computational cost and numerical stability must be balanced against limited observational information for boundary dynamics.\par

Several algorithms have recently been developed to reconstruct FSI from single-phase data. For instance, Kontogiannis et al. \cite{Kontogiannis2022} reported a Bayesian method for joint flow reconstruction and boundary segmentation from noisy MRV data. They modeled the 2D domain boundary using a signed distance function and employed an adjoint-based solver to optimize the inflow condition and boundary shape, subject to a hard constraint on the governing equations. Their approach yielded accurate reconstructions for steady 2D flows with fixed boundaries. Karnakov et al. \cite{Karnakov2024} used optimizing a discrete loss (ODIL), a grid-based inverse solver that minimizes a composite loss function with soft constraints on the measurements and governing equations. ODIL minimizes the loss with a Newton solver and can recover flow and structure states from sparse velocity data, inheriting the convergence and stability properties of traditional solvers in some cases. For FSI problems, the ODIL loss penalizes residuals from the discretized Navier--Stokes equations and a no-slip boundary condition, with additional regularization applied to the level set representation of the surface. However, the method struggles with non-convex or geometrically complex shapes, tending to approximate them as simpler convex forms, and its computational cost grows rapidly with domain size and time in large-scale 3D or unsteady problems. More recently, Buhendwa et al. \cite{Buhendwa2024} extended ODIL by integrating it into the JAX-Fluids solver to infer 3D obstacle shapes and flow fields in steady supersonic flows. By using explicit shock-capturing schemes and level set representations of the boundary, they achieved accurate shape reconstructions even in shock-dominated regimes. While these methods provide a strong foundation for FSI reconstruction, dynamic problems involving complex geometries remain challenging due to limitations in computational efficiency, numerical stability, and the observability (or lack thereof) of unsteady boundary dynamics.\par

Physics-informed neural networks (PINNs) \cite{Raissi2019a} are a machine learning technique that may help address key limitations of shape inference in DA for FSI. PINNs use one or more neural networks, often termed coordinate neural networks, to map space--time coordinates to the corresponding flow or structure fields. Physical consistency is enforced in a weak form by incorporating residuals of the governing equations into the loss function. A growing body of work has applied PINNs to inverse problems in FSI. Raissi et al.~\cite{Raissi2019b} were among the first to do this, applying PINNs to vortex-induced vibrations of a rigid cylinder elastically mounted in a uniform 2D flow. Using scattered observations, including an advected scalar concentration and structural displacements, they reconstructed the flow field and inferred the drag and lift forces acting on the cylinder. Building on this work, Kharazmi et al.~\cite{Kharazmi2021} extended PINNs to scenarios involving flexible cylinders, showing that PINNs can accommodate deforming structures in unsteady flows. Tang et al.~\cite{Tang2022} introduced transfer learning, wherein a pre-trained model from a related configuration is fine-tuned for a specific vortex-induced vibration case to reduce training cost.\par

More recent formulations have incorporated moving-boundary constraints for inverse FSI problems. Zhu et al.~\cite{Zhu2024} enforced moving-wall conditions within a PINN framework for incompressible flows with prescribed wall motion. Sundar et al.~\cite{Sundar2024} developed a boundary-aware PINN to recover pressure fields from coarse velocity data for unsteady flows past plunging foils. Calicchia et al.~\cite{Calicchia2023} reconstructed pressure fields near a swimming fish from planar PIV data by imposing interface constraints using measured body kinematics, enabling estimation of hydrodynamic forces from fluid-phase measurements. Wang et al.~\cite{Wang2025} proposed a dual-network framework that couples a boundary-tracking network with a flow network and reconstructed pressure fields for complex moving boundaries (e.g., swimming fish and beating left ventricles) using observed boundary trajectories and PIV velocity measurements. To address inverse problems with unknown geometries, Zhu et al.~\cite{Zhu2025} introduced a volume-distributed PINN with a learnable body-fraction field to infer hidden rigid boundaries while reconstructing the surrounding flow field from sparse velocity observations.\par

Although reconstruction algorithms that use two-phase data have shown promise for joint flow reconstruction and shape inference, they face significant practical hurdles. Simultaneously measuring both the fluid and structural phases is often costly, complex, and, in some cases, infeasible (e.g., blood flow through a cerebral aneurysm). In contrast, existing algorithms that rely solely on single-phase measurements can be less well suited for dynamic shape inference, in part because accurately reconstructing an evolving fluid--structure interface from incomplete data remains challenging. Consequently, there is a need for algorithms that can reconstruct FSI using sparse, noisy measurements from a single phase, particularly from the fluid phase alone. Ideally, such methods would enable inference of quantities such as boundary kinematics, material parameters, and dense flow fields (e.g., velocity and pressure) without requiring simultaneous two-phase measurements.\par

Although reconstruction algorithms that use two-phase data have shown promise for joint flow reconstruction and shape inference, they face significant practical hurdles. Simultaneously measuring both the fluid and structural phases is often costly, complex, and, in some cases, infeasible (e.g., blood flow through a cerebral aneurysm). In contrast, existing algorithms that rely solely on single-phase measurements can be less well suited for dynamic shape inference, in part because accurately reconstructing an evolving fluid--structure interface from incomplete data remains challenging. Consequently, there is a need for algorithms that can reconstruct FSI using sparse, noisy measurements from a single phase, particularly from the fluid phase alone. Ideally, such methods would enable inference of quantities such as boundary kinematics, material parameters, and dense flow fields (e.g., velocity and pressure) without requiring simultaneous two-phase measurements.\par

We present a DA framework for FSI that reconstructs fluid velocity and pressure fields, along with the structural response, from LPT data. While the framework can incorporate multi-modal measurements, its ability to operate using only single-phase data could potentially reduce experimental cost and complexity. The flow field and structure are represented using coordinate neural networks, with the structure parameterized using either physics-based or data-driven deformation modes. A specialized sampling strategy, implemented via adaptive Monte Carlo integration, is used to evaluate the loss terms over unsteady flow and structure domains. LPT data are embedded as a hard constraint through kinematics-constrained particle tracks, enabling physics-enhanced refinement of the tracks; the fluid governing equations are weakly enforced through an explicit physics loss. Structural motion is inferred by imposing a moving-wall condition at the fluid--solid interface and solving for the modal coefficients. The remainder of the paper is organized as follows: Sec.~\ref{sec: method} describes the algorithm and Sec.~\ref{sec: sampling} outlines the sampling scheme; Sec.~\ref{sec: cases} defines the test cases, Sec.~\ref{sec: results} presents results, and Sec.~\ref{sec: conclusions} concludes the manuscript.\par

%%% Methodology %%%
\section{Neural Inversion for Fluid--Structure Interactions}
\label{sec: method}
The objective of our DA algorithm is to reconstruct the fluid velocity field, pressure field, and structural response in an FSI system from LPT data (i.e., particle tracks). The reconstruction is constrained to be consistent with the fluid governing equations and the fluid--solid interface conditions. Our solver is based on parallel PINNs, with dedicated networks for the fluid and solid phases, together with particle models that embed advection kinematics as a hard constraint. This section introduces the overall framework in Sec.~\ref{sec: method: architecture}, with detailed descriptions of the flow-field, solid-surface, and particle-motion models in Secs.~\ref{sec: method: model: flow}--\ref{sec: method: model: KCT}.\par

% Architecture
\subsection{Data Assimilation Framework}
\label{sec: method: architecture}
\begin{figure}[htb!]
    \centering
    \includegraphics[width=\textwidth]{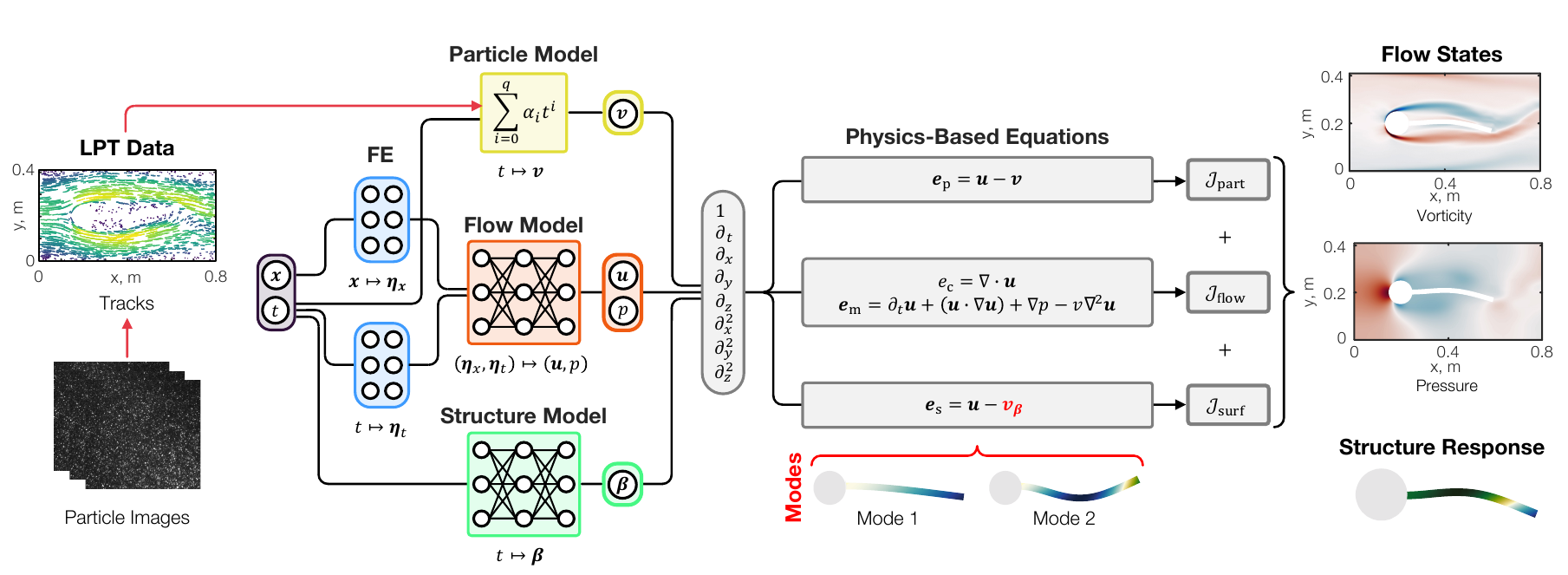}
    \caption{Architecture for FSI reconstruction. PINNs represent the flow and structure. Particle trajectories are enforced as hard constraints in the particle models, while soft constraints enforce the governing equations and the moving-wall boundary condition.}
    \label{fig: schematic}
\end{figure}

Our FSI DA framework consists of three main components: a neural flow model, a neural--modal surface model, and kinematics-constrained models of the particle tracks. These components are jointly trained by minimizing a composite physics-based loss. Figure~\ref{fig: schematic} illustrates the relationships among the models and their associated losses. The next two subsections describe the models and the corresponding loss terms.\par

% Models
\subsubsection{Taxonomy of Models}
\label{sec: method: architecture: models}
The fluid model, $\mathsf{F}$, is a ``fluid PINN'' constructed from one or more coordinate neural networks that map space--time inputs to flow-field quantities. For an incompressible flow, this PINN is represented as
\begin{equation}
    \label{equ: flow states}
    \mathsf{F} : \Omega \rightarrow \mathbb{R}^{d+1}, \quad
    (\boldsymbol{x}, t) \mapsto \left(\boldsymbol{u}, p\right),
\end{equation}
where $\boldsymbol{x} \in \mathcal{V}_t$ denotes spatial coordinates in the time-dependent flow domain $\mathcal{V}_t$, $t \in \mathcal{T}$ is time within the measurement interval $\mathcal{T}$, $\boldsymbol{u}$ and $p$ are the fluid velocity and pressure, respectively, and $d = 2$ or $3$ is the number of spatial dimensions. Additional quantities (e.g., density or total energy) can be included in the outputs of $\mathsf{F}$ when needed. The space--time domain $\Omega$ is the union of unsteady spatial domains over the measurement horizon,
\begin{equation}
    \Omega = \bigcup_{t \in \mathcal{T}} \mathcal{V}_t \times \{t\} \subset \mathbb{R}^d \times \mathbb{R}.
\end{equation}
Details of the fluid model architecture are provided in Sec.~\ref{sec: method: model: flow}.\par

The surface model, $\mathsf{S}$, represents the fluid--structure interface as a continuous manifold embedded in $\mathbb{R}^d$ (a 2D surface for $d = 3$, or a 1D curve for $d = 2$) using a set of discrete mesh points. The coordinates of these points at time $t$ are collected in the vector $\boldsymbol{p}\in \partial\mathcal{V}_t$, where each entry corresponds to a material point on the interface. The instantaneous surface position is approximated as the sum of a base surface, $\overline{\boldsymbol{p}}$, and a linear combination of $n_\mathrm{m}$ deflection\slash bending modes, $\boldsymbol{\Phi} = (\boldsymbol{\phi}_1, \dots, \boldsymbol{\phi}_{n_\mathrm{m}})$, introduced formally in Sec.~\ref{sec: method: model: structure}. The time-dependent modal coefficients are stored in $\boldsymbol{\beta} \in  \mathbb{R}^{n_\mathrm{m}}$, yielding the linear surface state
\begin{equation}
    \label{equ: modal structure}
    \boldsymbol{p} \approx \overline{\boldsymbol{p}} + \boldsymbol{\Phi}\,\boldsymbol{\beta}.
\end{equation}
The surface response is parameterized by a ``solid PINN,'' which maps time to the modal coefficients,
\begin{equation}
    \label{equ: structure states}
    \mathsf{S} : \mathcal{T} \rightarrow \mathbb{R}^{n_\mathrm{m}}, \quad
    t \mapsto \boldsymbol{\beta}.
\end{equation}
Differentiating $\boldsymbol{\beta}$ with respect to time yields the surface velocity.\par

Lastly, measurements of particle positions are grouped into discrete tracks, each containing a time-ordered sequence of positions for a single particle. The motion of the $k$th particle is described by a dedicated model, $\sdx{\mathsf{P}}[k]$, which maps time to that particle's velocity,
\begin{equation}
    \label{equ: particle tracks}
    \sdx{\mathsf{P}}[k] : \sdx{\mathcal{T}}[k] \rightarrow \mathbb{R}^d, \quad
    t \mapsto \sdx{\boldsymbol{v}}[k] \quad\text{for}\quad k \in \{1, 2, \dots, n_\mathrm{p}\},
\end{equation}
where $\sdx{\boldsymbol{v}}[k]$ is the predicted velocity of the $k$th particle at time $t$, $\sdx{\mathcal{T}}[k] \subseteq \mathcal{T}$ is the time interval covered by the track, and $n_\mathrm{p}$ is the total number of tracks. For the model to be physically admissible, its output must satisfy the kinematic advection constraint: when integrated in time, the predicted velocity must reproduce the tracked particle positions. Section~\ref{sec: method: model: KCT} describes how this constraint is embedded directly into the structure of $\sdx{\mathsf{P}}[k]$.\par

% Loss terms
\subsubsection{Loss Terms}
\label{sec: method: architecture: losses}
The flow, structure, and particle states, represented by $\mathsf{F}$, $\mathsf{S}$, and $\sdx{\mathsf{P}}[k]$, must satisfy the fluid governing equations, adhere to the moving-wall condition, and remain consistent with the measurements. As noted above, agreement of $\sdx{\mathsf{P}}[k]$ with the LPT data can be strictly enforced by its formulation. The flow equations and interface constraints, by contrast, are weakly enforced by minimizing a composite loss function that includes residuals from the governing equations and boundary conditions. The loss balances data fidelity with physics-based constraints:
\begin{equation}
    \label{equ: objective loss}
    \mathscr{J}_\mathrm{total} =
    \gamma_1 \,\mathscr{J}_\mathrm{flow} +
    \gamma_2 \,\mathscr{J}_\mathrm{surf} +
    \gamma_3 \,\mathscr{J}_\mathrm{part} +
    \gamma_4 \,\mathscr{J}_\mathrm{data},
\end{equation}
where $\gamma_i$ are weighting coefficients chosen to balance the contributions of the loss terms. The components $\mathscr{J}_\mathrm{flow}$, $\mathscr{J}_\mathrm{surf}$, $\mathscr{J}_\mathrm{part}$, and $\mathscr{J}_\mathrm{data}$ encode the fluid governing equations, interface condition, particle transport physics, and data fidelity, respectively.\par

The flow physics loss is defined as
\begin{equation}
    \label{equ: flow loss} 
    \mathscr{J}_\mathrm{flow} =
    \frac{\mathrm{dim}(\boldsymbol{e}_\mathrm{f})^{-1}}{\left| \Omega \right|}
    \iint_\Omega \left\lVert \boldsymbol{e}_\mathrm{f}
    \right\rVert_2^2 \mathrm{d}\boldsymbol{x} \,\mathrm{d}t,
\end{equation}
where $\boldsymbol{e}_\mathrm{f} \in \mathbb{R}^{d+1}$ contains the residuals of the incompressible continuity and Navier--Stokes equations,
\begin{subequations}
    \label{equ: fluid equations}
    \begin{align}
        e_\mathrm{c} &= \nabla \cdot \boldsymbol{u}, \label{equ: continuity} \\
        \boldsymbol{e}_\mathrm{m} &= \frac{\partial \boldsymbol{u}}{\partial t} + \left(\boldsymbol{u} \cdot \nabla\right)\mathopen{} \boldsymbol{u} + \nabla p - \nu \nabla^2 \boldsymbol{u}, \label{equ: momentum}
    \end{align}
\end{subequations}
where $\nu$ is the kinematic viscosity and $\boldsymbol{e}_\mathrm{f} = (e_\mathrm{c}, \boldsymbol{e}_\mathrm{m})$. Integration over $\Omega$ is interpreted in the Lebesgue sense, with $\mathrm{d}\boldsymbol{x}$ and $\mathrm{d}t$ indicating the standard Lebesgue measures $\mathrm{d}\mu(\boldsymbol{x})$ and $\mathrm{d}\mu(t)$. That is, $\mu(\mathcal{V}_t)$ gives the spatial volume of the flow domain at time $t$ and $\mu(\mathcal{T})$ gives the total duration of the reconstruction interval. This convention is used throughout the paper.\par

The boundary loss is defined as
\begin{equation}
    \label{equ: boundary loss}
    \mathscr{J}_\mathrm{surf} =
    \frac{\mathrm{dim}(\boldsymbol{e}_\mathrm{s})^{-1}}{|\partial\Omega|} \iint_{\partial\Omega}
    \left\lVert \boldsymbol{e}_\mathrm{s} \right\rVert_2^2 \mathrm{d}\boldsymbol{x} \,\mathrm{d}t,
\end{equation}
where $\partial\Omega$ is the space--time surface formed by the union of fluid--structure interfaces over $\mathcal{T}$,
\begin{equation}
    \partial\Omega = \bigcup_{t \in \mathcal{T}} \partial\mathcal{V}_t \times \{t\} \subset \mathbb{R}^d \times \mathbb{R},
\end{equation}
with $\partial\mathcal{V}_t$ being the portion of the flow domain boundary corresponding to the interface at time $t$. The residual vector $\boldsymbol{e}_\mathrm{s}$ quantifies the mismatch between the fluid velocity predicted by $\mathsf{F}$ and the surface velocity predicted by $\mathsf{S}$, evaluated at points $(\boldsymbol{x}, t) \in \partial\Omega$. This term weakly enforces a no-slip condition along the moving solid surface.\par

A particle physics loss is introduced to couple the particle model, which encodes the LPT measurements, to the flow and structure models,
\begin{equation}
    \label{equ: particle loss}
    \mathscr{J}_\mathrm{part} =
    \frac{\mathrm{dim}\mathopen{} \left( \sdx{\boldsymbol{e}}[k][\mathrm{p}] \right)^{-1}}{n_\mathrm{p}}
    \sum_{k=1}^{n_\mathrm{p}} \left(\frac{1}{\left|\sdx{\mathcal{T}}[k]\right|} \int_{\sdx{\mathcal{T}}[k]} \left\lVert \sdx{\boldsymbol{e}}[k][\mathrm{p}] \right\rVert_2^2 \mathrm{d}t\right),
\end{equation}
where $\sdx{\boldsymbol{e}}[k][\mathrm{p}]$ is the residual vector associated with the governing equation for the $k$th particle. For tracer particles that passively follow the flow, the particle velocity is assumed to match the fluid velocity, i.e., $\boldsymbol{v} = \boldsymbol{u}$, yielding the residual
\begin{equation}
    \sdx{\boldsymbol{e}}[k] =
    \boldsymbol{u}\mathopen{} \left[\sdx{\boldsymbol{x}}[k](t), t\right] -
    \sdx{\boldsymbol{v}}[k]\mathopen{} \left(t\right).
\end{equation}
Here, $\sdx{\boldsymbol{v}}[k]$ is the particle velocity predicted by model $\sdx{\mathsf{P}}[k]$ and $\boldsymbol{u}$ is the fluid velocity from $\mathsf{F}$, evaluated at the corresponding particle position determined from $\sdx{\mathsf{P}}[k]$. This term couples the particle and fluid states. The framework can also accommodate positional uncertainty and non-ideal tracer behavior (e.g., inertial transport effects), as demonstrated in Refs.~\cite{Zhou2023b, Zhou2025}.\par

To account for localization uncertainty in measured particle positions, we treat the particle positions at measurement times as trainable variables alongside the weights and biases of the neural networks. This allows trajectories to be adjusted during reconstruction. To discourage drift away from the measured LPT data, the data fidelity term penalizes deviations with a weight determined by the assumed localization uncertainty. The data loss is formulated as
\begin{equation}
    \label{equ: data loss}
    \mathscr{J}_\mathrm{data} =
    \left(\frac{1}{d \, n_\mathrm{p}}
    \sum_{k=1}^{n_\mathrm{p}} \frac{1}{n_k}
    \sum_{i=1}^{n_k} \left\lVert \sdx{\boldsymbol{x}}[k][i] - \sdx{\hat{\boldsymbol{x}}}[k][i] \right\rVert_{\boldsymbol{L}}^2 - 1
    \right)^2,
\end{equation}
where $\sdx{\boldsymbol{x}}[k][i]$ and $\sdx{\hat{\boldsymbol{x}}}[k][i]$ are the measured and estimated (trainable) positions of the $k$th particle at time $t_i$, with $n_\mathrm{p}$ total tracks and $n_k$ samples in the $k$th track. The norm is the matrix-weighted Mahalanobis norm,
\begin{equation}
    \label{equ: Mahalanobis norm}
    \lVert \Delta\boldsymbol{x} \rVert_{\boldsymbol{L}}^2 = \Delta\boldsymbol{x}^\top \boldsymbol{L}^\top \boldsymbol{L} \, \Delta\boldsymbol{x},
    \quad \text{where} \quad
    \boldsymbol{L}^\top \boldsymbol{L} = \boldsymbol{\Gamma}^{-1},
\end{equation}
and $\boldsymbol{\Gamma}$ is the covariance matrix of the positional uncertainty. For independent, centered Gaussian localization errors with standard deviations $\sigma_j$ along each axis, $\boldsymbol{\Gamma} = \mathrm{diag}(\sigma_1^2, \ldots, \sigma_d^2)$. \par

The estimated positions $\sdx{\hat{\boldsymbol{x}}}[k][i]$ are embedded in the kinematics-constrained track model through the displacement vector $\boldsymbol{\delta}$ (Sec.~\ref{sec: method: model: KCT}). For data with appreciable noise, treating $\sdx{\hat{\boldsymbol{x}}}[k][i]$ as trainable parameters enables physics-informed trajectory refinement. The data loss is balanced by the physics and boundary losses, $\mathscr{J}_\mathrm{flow}$ and $\mathscr{J}_\mathrm{part}$, which help to constrain the track geometry. This can improve the particle position estimates and, in turn, the reconstructed flow fields. When the localization uncertainty is negligible, $\sdx{\hat{\boldsymbol{x}}}[k][i]$ may be fixed at the measured values, in which case $\mathscr{J}_\mathrm{data} = 0$.\par

The integrals in Eqs.~\eqref{equ: flow loss}, \eqref{equ: boundary loss}, and \eqref{equ: particle loss} cannot be evaluated in closed form, but they may be approximated using Monte Carlo sampling. Accurate integration requires a sampling scheme that accounts for the evolving spatial domain $\mathcal{V}_t$ and the moving boundary $\partial \mathcal{V}_t$, which are estimated using the transient surface shape from $\mathsf{S}$. Our sampling strategy and associated numerical considerations are described in Sec.~\ref{sec: sampling}. The data loss is particularly relevant for single-camera LPT modalities such as plenoptic imaging or digital in-line holography~\cite{Zhou2025}, where localization errors can be substantial. Joint minimization of the composite loss produces particle positions and velocities that are statistically consistent with the experimental data, while yielding flow and structure states that approximately satisfy the governing equations and enforce the no-slip condition at the moving interface.\par

% Flow
\subsection{Flow Model}
\label{sec: method: model: flow}
Flow states are represented by one or more coordinate neural networks. The generic architecture of these deep, feed-forward networks, $\mathsf{N} : \sdx{\boldsymbol{z}}[0] \mapsto \sdx{\boldsymbol{z}}[n_\mathrm{L} + 1]$, consists of an input layer, an output layer, and a sequence of $n_\mathrm{L}$ hidden layers,
\begin{subequations}
    \label{equ: architecture}
    \begin{gather}
        \sdx{\boldsymbol{z}}[n_\mathrm{L} + 1] =
        \mathsf{N}\mathopen{} \left(\sdx{\boldsymbol{z}}[0]\right) =
        \sdx{\boldsymbol{W}}[n_\mathrm{L}+1] \left[\sdx{\mathsf{L}}[n_\mathrm{L}] \circ \sdx{\mathsf{L}}[n_\mathrm{L}-1] \circ \dots \circ \sdx{\mathsf{L}}[2] \circ \mathsf{G}\mathopen{}\left(\sdx{\boldsymbol{z}}[0]\right)\right] + \sdx{\boldsymbol{b}}[n_\mathrm{L}+1],
        \intertext{with hidden layers $\sdx{\mathsf{L}}[l] : \sdx{\boldsymbol{z}}[l-1] \mapsto \sdx{\boldsymbol{z}}[l]$ given by}
        \sdx{\boldsymbol{z}}[l] =
        \sdx{\mathsf{L}}[l]\mathopen{} \left(\sdx{\boldsymbol{z}}[l-1]\right) =
        \mathrm{swish}\mathopen{} \left(\sdx{\boldsymbol{W}}[l]\sdx{\boldsymbol{z}}[l-1] + \sdx{\boldsymbol{b}}[l]\right) \quad\text{for}\quad l \in \{2, 3, \dots, n_\mathrm{L}\}.
    \end{gather}
\end{subequations}
Here, $\sdx{\boldsymbol{z}}[l]$ contains the neuron values in the $l$th layer, $\sdx{\boldsymbol{W}}[l]$ and $\sdx{\boldsymbol{b}}[l]$ are the corresponding weight matrix and bias vector, and $\circ$ denotes function composition. Swish activations are used in all hidden layers, defined element-wise as
\begin{equation}
    \label{equ: swish}
    \mathrm{swish}(z) = \frac{z \,\exp(z)}{1 + \exp(z)}.
\end{equation}
To mitigate the spectral bias of gradient-based training \cite{Wang2021}, the first hidden layer, $\sdx{\mathsf{L}}[1]$, is replaced with a Fourier encoding \cite{Tancik2020},
\begin{equation}
    \label{equ: method:FE}
    \sdx{\boldsymbol{z}}[1] =
    \mathsf{G}\mathopen{} \left(\sdx{\boldsymbol{z}}[0]\right) = \left[\sin\mathopen{}\left(2\pi \boldsymbol{f}_1 \cdot \sdx{\boldsymbol{z}}[0]\right), \,\cos\mathopen{}\left(2\pi \boldsymbol{f}_1 \cdot \sdx{\boldsymbol{z}}[0]\right), \dots, \,\sin\mathopen{}\left(2\pi \boldsymbol{f}_{n_\mathrm{F}} \cdot \sdx{\boldsymbol{z}}[0]\right), \,\cos\mathopen{}\left(2\pi \boldsymbol{f}_{n_\mathrm{F}} \cdot \sdx{\boldsymbol{z}}[0]\right)\right],
\end{equation}
where $\boldsymbol{f}_j$ is a vector of random frequencies assigned to each input dimension and $n_\mathrm{F}$ is the number of Fourier features. The frequencies are drawn from a zero-mean Gaussian distribution before training and fixed thereafter. The standard deviation of the frequency distribution controls the spectral bandwidth of functions that the network can represent efficiently \cite{Tancik2020}.\par

A single network can learn vector-valued functions whose components have similar spectral content, but its efficiency may decrease when those components exhibit distinct spectral characteristics. In turbulent flows, for example, velocity and pressure often have different wavenumber scalings. Using separate networks for $\boldsymbol{u}$ and $p$ can therefore improve reconstruction accuracy at a fixed model size. We thus use dedicated networks for velocity and pressure, $\mathsf{F}_{\boldsymbol{u}} : (\boldsymbol{x}, t) \mapsto \boldsymbol{u}$ and $\mathsf{F}_p : (\boldsymbol{x}, t) \mapsto p$, following the architecture in Eq.~\eqref{equ: architecture}. Automatic differentiation of these networks is used to evaluate the residuals in Eq.~\eqref{equ: fluid equations}.\par

% Structure
\subsection{Structure Model}
\label{sec: method: model: structure}
Representing time-varying geometries is a central challenge in FSI flow reconstruction and inverse design. The fluid--solid interface $\partial\mathcal{V}_t$ may undergo complex deformations, and our goal is to infer its shape over time from sparse measurements such as particle tracks or MRV data. Because both the surface and surrounding flow field are, in principle, infinite-dimensional, reconstructing an arbitrary surface from sparse observations is ill posed. To regularize the problem, we adopt a low-dimensional modal representation of the interface, approximating the instantaneous surface as a linear combination of deformation modes obtained from (i)~physics-based eigenmode analysis or (ii)~data-driven analysis of surface measurements. This linear expansion is most appropriate for smooth, globally coherent deformations (e.g., bending and twisting in slender structures), whereas localized high-curvature features such as wrinkling or folding may require many more modes or a nonlinear representation.\par

We treat the evolving interface $\partial\mathcal{V}_t$ as a smooth $(d - 1)$-dimensional manifold embedded in $\mathbb{R}^d$ \cite{Lee2003, Ma2012}. Due to curvature and potentially complex topology, a global chart is generally unavailable. Instead, we define an atlas of local charts,
\begin{equation}
    \mathcal{A} = \left\{\left(\sdx{\Theta}[i], \sdx{\mathsf{M}}[i]\right) \mid i \in \mathcal{I}\right\},
\end{equation}
where $\sdx{\Theta}[i] \subset \mathbb{R}^{d-1}$ is a local coordinate domain and $\sdx{\mathsf{M}}[i] : \sdx{\Theta}[i] \times \mathcal{T} \rightarrow \partial\mathcal{V}_t$ defines a smooth, time-dependent embedding of a surface patch. The full surface is given by the union
\begin{equation}
    \partial \mathcal{V}_t = \bigcup_{i \in \mathcal{I}}
    \left\{\sdx{\mathsf{M}}[i](\boldsymbol{\theta}, t) \mid \boldsymbol{\theta} \in \sdx{\Theta}[i]\right\}.
\end{equation}
Each chart is assumed to be regular and locally injective, ensuring that the surface is non-degenerate and free of self-intersections within each patch. We also assume the existence of a local inverse on each patch, $\sdx{\mathsf{M}}[i]{^{-1}}$, defined on $\sdx{\mathsf{M}}[i](\sdx{\Theta}[i]\times\mathcal{T}) \subset \partial\Omega$, which maps a point on the interface back to local coordinates, $(\boldsymbol{x},t) \mapsto \boldsymbol{\theta}$. This provides a coordinate system anchored to material points on the surface, though it does not accommodate geometric singularities or topological changes such as merging or tearing.\par

To construct a modal basis using a proper orthogonal decomposition (POD), we define a global parameter space by the disjoint union
\begin{equation}
    \Theta =
    \bigsqcup_{i \in \mathcal{I}} \sdx{\Theta}[i],
\end{equation}
and we assume smooth blending through a partition of unity \cite{Melenk1996}. The mapping $\mathsf{M} : \Theta \times \mathcal{T} \rightarrow \mathbb{R}^d$ provides global surface coordinates, enabling us to define mean and fluctuating positions on the surface as
\begin{subequations}
    \begin{align}
        \overline{\boldsymbol{x}}(\boldsymbol{\theta}) &=
        \frac{1}{|\mathcal{T}|} \int_\mathcal{T} \mathsf{M}(\boldsymbol{\theta}, t) \,\mathrm{d}t, \\
        \delta\boldsymbol{x}(\boldsymbol{\theta}, t) &=
        \mathsf{M}(\boldsymbol{\theta}, t) - \overline{\boldsymbol{x}}(\boldsymbol{\theta}).
    \end{align}
\end{subequations}
To obtain coherent deformation modes, we define the vector-valued surface displacement covariance operator \cite{Berkooz1993, Taira2017}. In kernel form, this operator is represented by
\begin{equation}
    \label{equ: structure:covariance_kernel}
    \boldsymbol{K}(\boldsymbol{\theta}, \boldsymbol{\theta}^\prime) =
    \frac{1}{\left| \mathcal{T} \right|} \int_\mathcal{T}
    \delta\boldsymbol{x}(\boldsymbol{\theta}, t)\,
    \delta\boldsymbol{x}(\boldsymbol{\theta}^\prime, t)^\top
    \,\mathrm{d}t,
\end{equation}
with $\boldsymbol{K} \in \mathbb{R}^{d \times d}$, and we solve the associated eigenproblem
\begin{equation}
    \label{equ: structure:POD_eigenproblem}
    \int_{\Theta} \boldsymbol{K}(\boldsymbol{\theta}, \boldsymbol{\theta}^\prime)\,
    \boldsymbol{\varphi}_j(\boldsymbol{\theta}^\prime) \,\mathrm{d}\boldsymbol{\theta}^\prime =
    \lambda_j \,\boldsymbol{\varphi}_j(\boldsymbol{\theta}),
\end{equation}
where the modes $\boldsymbol{\varphi}_j : \Theta \rightarrow \mathbb{R}^d$ are orthonormal with respect to the $L^2(\Theta)$ inner product,
\begin{equation}
    \label{equ: structure:orthonormality}
    \int_{\Theta} \boldsymbol{\varphi}_j(\boldsymbol{\theta})
    \cdot \boldsymbol{\varphi}_k(\boldsymbol{\theta})
    \,\mathrm{d}\boldsymbol{\theta} = \delta_{jk}.
\end{equation}
We then define a truncated modal expansion
\begin{equation}
    \label{equ: structure:modal_expansion_continuous}
    \delta\boldsymbol{x}(\boldsymbol{\theta}, t) \approx
    \sum_{j=1}^{n_\mathrm{m}} \beta_j(t) \,\boldsymbol{\varphi}_j(\boldsymbol{\theta}),
\end{equation}
with coefficients
\begin{equation}
    \label{equ: structure:modal_coefficients}
    \beta_j(t) = \int_{\Theta} \delta\boldsymbol{x}(\boldsymbol{\theta}, t) \cdot \boldsymbol{\varphi}_j(\boldsymbol{\theta}) \,\mathrm{d}\boldsymbol{\theta}.
\end{equation}
Given an appropriate choice of $n_\mathrm{m}$, this continuous representation provides a convenient reduced-order description of surface kinematics and motivates our discrete implementation.\par

In practice, both the base surface and deformation modes are evaluated at a discrete set of nodal positions $\{\sdx{\boldsymbol{\theta}}[1], \dots, \sdx{\boldsymbol{\theta}}[n_\mathrm{s}]\}$, which define a material mesh over the deforming boundary $\partial \mathcal{V}_t$. The corresponding physical positions $\sdx{\boldsymbol{x}}[k] \in \mathbb{R}^d$ are stacked into a global vector $\boldsymbol{p} \in \mathbb{R}^{d n_\mathrm{s}}$. The (time-averaged) base surface is given by $\overline{p}_{(k - 1)d + \ell} = \overline{x}_\ell(\sdx{\boldsymbol{\theta}}[k])$, and the mode matrix is defined by
\begin{equation}
    \label{equ: structure:mode_matrix}
    \Phi_{(k-1)d+\ell,\, j} = \big[\boldsymbol{\varphi}_j(\sdx{\boldsymbol{\theta}}[k])\big]_\ell,
    \quad k \in \{1,\dots,n_\mathrm{s}\},\ \ell \in \{1,\dots,d\}.
\end{equation}
Surface motion is governed by the solid PINN $\mathsf{S}$, configured according to Eq.~\eqref{equ: architecture}, which outputs time-dependent modal coefficients $\boldsymbol{\beta}$. The resulting surface at time $t$ is approximated as
\begin{equation}
    \label{equ: structure:modal_surface_discrete}
    \boldsymbol{p}(t) = \overline{\boldsymbol{p}} + \boldsymbol{\Phi} \boldsymbol{\beta}(t),
\end{equation}
with the corresponding surface velocity obtained by automatic differentiation of $\mathsf{S}$ with respect to $t$,
\begin{equation}
    \label{equ: surface velocity}
    \frac{\mathrm{d}\boldsymbol{p}}{\mathrm{d}t} = \boldsymbol{\Phi} \frac{\mathrm{d}\boldsymbol{\beta}}{\mathrm{d}t}.
\end{equation}
This velocity provides the kinematic boundary condition needed to evaluate the surface loss term in Eq.~\eqref{equ: boundary loss}. The modal basis $\boldsymbol{\Phi}$ may be constructed a priori through finite element analysis using estimated material properties, or extracted a posteriori by applying a modal decomposition to time-resolved surface tracking data (e.g., in multi-modal FSI experiments or from asynchronous surface measurements). Importantly, these structural data need not be acquired simultaneously with the flow measurements. Once established, the modal basis constrains the admissible deformation space while the specific temporal evolution $\boldsymbol{\beta}(t)$ is inferred solely from the particle tracks. Both strategies are demonstrated synthetically in Sec.~\ref{sec: results}.\par

% Light tightening + a few formal/notation fixes for the Particle Track Model.
% Source: :contentReference[oaicite:0]{index=0}

\subsection{Particle Track Model}
\label{sec: method: model: KCT}
We define a kinematics-constrained model $\sdx{\mathsf{P}}[k]$ for each particle track, which maps time to the particle velocity while satisfying the advection constraint,
\begin{equation}
    \label{equ: particle advection}
    \frac{\mathrm{d} \sdx{\boldsymbol{x}}[k]}{\mathrm{d}t} = \sdx{\boldsymbol{v}}[k]
    \;\Longrightarrow\;
    \sdx{\boldsymbol{x}}[k][j+1] - \sdx{\boldsymbol{x}}[k][j] =
    \int_{t_{j}}^{t_{j+1}} \sdx{\boldsymbol{v}}[k]\mathopen{}\left(t\right) \mathrm{d}t,
\end{equation}
where $\sdx{\boldsymbol{x}}[k]$ and $\sdx{\boldsymbol{v}}[k]$ are the $k$th particle's position and velocity, and $j$ indexes measurement times $t_j$. Each track comprises a sequence of $n_k$ positions $\sdx{\boldsymbol{x}}[k][j]$ recorded at discrete times $t_j$. To enable efficient gradient-based optimization, we embed the integral constraints into the model using the theory of functional connections \cite{Leake2022}, which reformulates the constrained problem in terms of unconstrained parameters. We adopt a projection--switch form for one component of the velocity,
\begin{equation}
    \label{equ: track polynomial}
    \mathsf{P}\mathopen{}\left(t\right) \equiv v\mathopen{}\left(t\right) =
    g\mathopen{}\left(t\right) + \sum_{j=1}^{n_\mathrm{c}} \eta_j \,\xi_j\mathopen{}\left(t\right),
\end{equation}
where $g$ is a free function with tunable parameters, and $\eta_j$ and $\xi_j$ are projection coefficients and switch functions that enforce the integral constraints over each interval $[t_{j-1}, t_j]$. The number of constraints is $n_\mathrm{c} = n_k - 1$. A separate copy of this model is used for each spatial dimension, and the track index $k$ is omitted here for brevity.\par

The free function is chosen as a $q$th-order polynomial,
\begin{equation}
    \label{equ: free function}
    g\mathopen{}\left(t\right) = \sum_{i = 0}^{q} \alpha_i \,t^i,
\end{equation}
with trainable coefficients $\alpha_i$. To enforce the advection constraint, we define the projection coefficients as
\begin{equation}
    \label{equ: projection coefficients}
    \eta_j = \left(x_j - x_{j-1}\right) - \int_{t_{j-1}}^{t_j} g\mathopen{}\left(t\right) \mathrm{d}t,
\end{equation}
which encodes the correction needed for $g$ to match the observed displacement over each interval. These corrections are applied using switch functions
\begin{equation}
    \label{equ: switch function}
    \xi_j\mathopen{}\left(t\right) = \sum_{i = 1}^{n_\mathrm{c}} h_i\mathopen{}\left(t\right) L_{ij},
\end{equation}
where $h_i(t) = t^{i-1}$ are monomial support functions and $L_{ij}$ are entries of a weight matrix $\boldsymbol{L}$. This matrix is chosen so that the switch functions integrate to unity over their corresponding intervals and vanish elsewhere, enforced by
\begin{equation}
    \label{equ: linear constraints}
    \boldsymbol{H}\boldsymbol{L} = \boldsymbol{I},
\end{equation}
where $\boldsymbol{H}$ has entries
\begin{equation}
    H_{ij} = \int_{t_{i-1}}^{t_i} h_j\mathopen{}\left(t\right) \mathrm{d}t.
\end{equation}
The monomial basis ensures that $\boldsymbol{H}$ is invertible, yielding a unique solution for $\boldsymbol{L}$.\par

Given $g$ and a particle track, a closed-form expression for $v$ follows from (1)~computing $\boldsymbol{\eta}$ via Eq.~\eqref{equ: projection coefficients} and (2)~solving Eq.~\eqref{equ: linear constraints} for $\boldsymbol{L}$ to specify $\boldsymbol{\xi}$. A compact matrix form is
\begin{equation}
    \label{equ: track matrix}
    v\mathopen{}\left(t\right) =
    \left[\boldsymbol{\alpha}^\top \left(\boldsymbol{I} - \boldsymbol{C}\widehat{\boldsymbol{L}}\right) + \boldsymbol{\delta}^\top \widehat{\boldsymbol{L}}\right]
    \boldsymbol{\tau}\mathopen{}\left(t\right),
\end{equation}
where $\boldsymbol{\tau}(t) = \{t^j \mid j = 0, \dots, p\}$ is a time basis vector, $\boldsymbol{\delta} = \{x_j - x_{j-1} \mid j = 1, \dots, n_\mathrm{c}\}$ is a displacement vector, and $\boldsymbol{C}$ is a support matrix with elements
\begin{equation}
     C_{ij} = i^{-1}\left(t_j^i - t_{j-1}^i\right).
\end{equation}
Here, $\widehat{\boldsymbol{L}} = [\boldsymbol{L}; \mathbf{0}]^\top$ is an $n_\mathrm{c} \times (p + 1)$ augmented weight matrix. This velocity model is differentiable and satisfies the advection constraint for any choice of coefficients $\boldsymbol{\alpha} \in \mathbb{R}^{q+1}$.\par

Integrating $v$ yields a continuous particle trajectory,
\begin{equation}
    \label{equ: continuous particle trajectory}
    x\mathopen{}\left(t\right) =
    \left[\boldsymbol{\alpha}^\top \left(\boldsymbol{I} - \boldsymbol{C}\widehat{\boldsymbol{L}}\right) + \boldsymbol{\delta}^\top \widehat{\boldsymbol{L}}\right]
    \widetilde{\boldsymbol{\tau}}\mathopen{} \left(t\right) + x_0,
\end{equation}
where $x_0$ is the initial particle position and $\widetilde{\boldsymbol{\tau}} (t) = \{t^{j+1}/(j+1) \mid j=0, \dots, p\}$. This formulation enables differentiable evaluation of particle positions and velocities at arbitrary times $t \in \sdx{\mathcal{T}}[k]$, and it is used to compute the particle loss in Eq.~\eqref{equ: particle loss}.\par

%%% Sampling %%%
\section{Approximating Loss Terms for Transient Domains}
\label{sec: sampling}
To train the fluid, structure, and particle models, we minimize a composite loss functional $\mathscr{J}_\mathrm{total}$ defined in terms of residuals over continuous space--time volumes and surfaces. These residuals arise from the fluid's governing equations, interface boundary conditions, and particle transport equations, and they are integrated over the transient fluid domain $\Omega$, the fluid--structure interface $\partial \Omega$, and the track intervals $\sdx{\mathcal{T}}[k]$. Explicit expressions for these loss terms are given in Eqs.~\eqref{equ: flow loss}, \eqref{equ: boundary loss}, and \eqref{equ: particle loss}. Since the integrals cannot be evaluated in closed form, we approximate them using Monte Carlo sampling,
\begin{equation}
    \mathscr{J} =
    \frac{1}{|\Omega|} \iint_{\Omega} \mathscr{L}\mathopen{} \left(\boldsymbol{x}, t\right) \mathrm{d}\boldsymbol{x} \,\mathrm{d}t \approx
    \frac{1}{N} \sum_{i=1}^N \mathscr{L}\mathopen{} \left(\sdx{\boldsymbol{x}}[i], \sdx{t}[i]\right),
\end{equation}
where $\mathscr{L}$ is a squared residual term and $(\sdx{\boldsymbol{x}}[i], \sdx{t}[i])$ are samples drawn from the relevant domain. Because the integrals are defined with respect to Lebesgue measures on $\Omega$, $\partial \Omega$, and $\sdx{\mathcal{T}}[k]$, we construct the sampling scheme to approximate uniform coverage with respect to those measures. Time samples are drawn uniformly from $\mathcal{T}$ or $\sdx{\mathcal{T}}[k]$, and spatial samples are allocated in proportion to the length (1D), area (2D), or volume (3D) of subregions of $\mathcal{V}_t$ or $\partial \mathcal{V}_t$. This ensures consistency between the sampling distribution and the continuous measure used in each loss term. Residuals ($\boldsymbol{e}_\mathrm{f}$, $\boldsymbol{e}_\mathrm{s}$, and $\sdx{\boldsymbol{e}}[k][\mathrm{p}]$) are evaluated using the fluid, structure, and particle models, and gradients of the total loss are computed by automatic differentiation and used in backpropagation.\par

The effectiveness of this approach depends on the sampling scheme \cite{Berrone2022, Mao2023, Wan2024, Taylor2025}. Poor sampling can introduce variance in the integral approximations, leading to noisy gradients and degraded convergence \cite{Molnar2025}. We therefore implement a sampling strategy that tracks the evolving geometry of the fluid and structure domains. In doing so, we aim to make each loss term a faithful approximation of its continuous counterpart and to promote convergence toward a weak-form solution of the governing equations and boundary conditions. We outline the sampling strategy for each term below with reference to the 2D and 3D FSI geometries tested in this work; complete case definitions are deferred to Sec.~\ref{sec: cases}.\par

% Fluid domain
\subsection{Fluid Domain Sampling}
\label{sec: sampling: fluid}
\begin{figure}[htb!]
    \vspace*{-0.25em}
    \centering
    \includegraphics[width=0.9\textwidth]{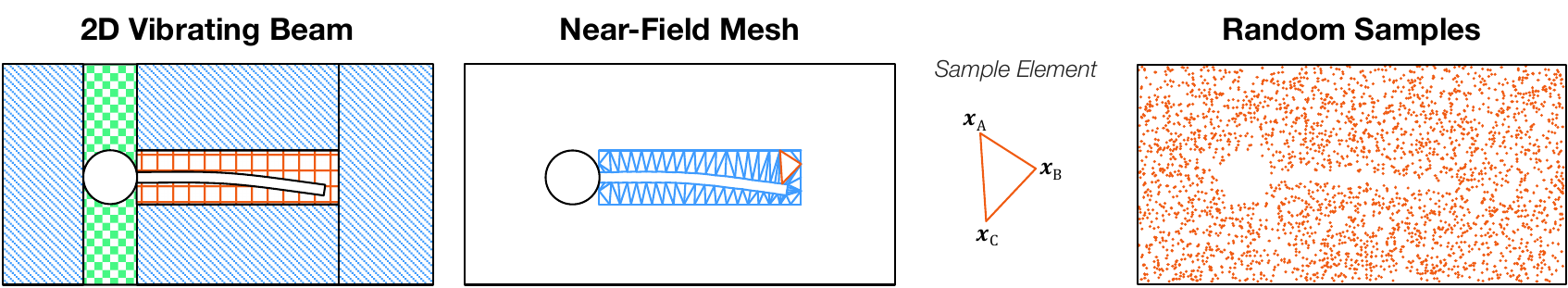}
    \vspace*{0.5em}
    \caption{Sampling strategy. Left: the domain is partitioned into far-field regions (blue), near-field regions around rigid structures (green), and a deforming near-field region around the compliant structure (red). Middle: the deforming region is discretized into elements anchored to surface nodes $\boldsymbol{p}$. Right: samples are drawn from each subregion in proportion to its area, yielding approximately uniform coverage of $\mathcal{V}_t$.}
    \label{fig: sampling}
\end{figure}

The fluid domain is partitioned into subregions based on proximity to solid boundaries and geometric complexity. As illustrated in Fig.~\ref{fig: sampling} for the 2D flapping plate case (Sec.~\ref{sec: cases: flow: 2D}), we define three subdomain types: (1)~fixed far-field rectangles (blue), (2)~rectangles containing a circular cutout around the rigid cylinder (green), and (3)~a deforming near-field region adjacent to the flapping plate, discretized by a triangular mesh (red). Sampling is performed at randomly selected time instants $t \in \mathcal{T}$ drawn uniformly over the time horizon. At each time, spatial locations $\boldsymbol{x} \in \mathcal{V}_t$ are sampled from each subregion in proportion to its area (2D) or volume (3D), yielding approximately uniform coverage of the time-dependent fluid domain. In axis-aligned rectangular regions, we use independent uniform distributions over each coordinate. For circular cutouts, samples are drawn from the bounding rectangle and points inside the cutout are rejected.\par

The deforming near-field region is handled via mesh-based sampling. At each sampled time, the surface mesh is updated using the predicted mode coefficients $\boldsymbol{\beta}$ from $\mathsf{S}$. In 2D, triangular elements are sampled with probability proportional to their area; in 3D, tetrahedra are weighted by instantaneous volume. Within each element, random points are generated using barycentric coordinates \cite{Hormann2017}. For a triangle with vertices $\boldsymbol{x}_\mathrm{A}$, $\boldsymbol{x}_\mathrm{B}$, and $\boldsymbol{x}_\mathrm{C}$ (Fig.~\ref{fig: sampling}), we set
\begin{gather*}
    \boldsymbol{x} =
    (1 - \sqrt{r_1}) \,\boldsymbol{x}_\mathrm{A} +
    \sqrt{r_1} (1 - r_2) \,\boldsymbol{x}_\mathrm{B} +
    \sqrt{r_1} r_2 \,\boldsymbol{x}_\mathrm{C},
    \intertext{where $r_1, r_2 \sim \mathcal{U}(0,1)$. The 3D analogue, with a fourth vertex $\boldsymbol{x}_\mathrm{D}$, is}
    \boldsymbol{x} =
    (1 - \sqrt[3]{r_1}) \,\boldsymbol{x}_\mathrm{A} +
    \sqrt[3]{r_1}(1 - \sqrt{r_2}) \,\boldsymbol{x}_\mathrm{B} +
    \sqrt[3]{r_1} \sqrt{r_2}(1 - r_3) \,\boldsymbol{x}_\mathrm{C} +
    \sqrt[3]{r_1} \sqrt{r_2} r_3 \,\boldsymbol{x}_\mathrm{D},
\end{gather*}
where $r_1, r_2, r_3 \sim \mathcal{U}(0,1)$. The mesh is used only to support rapid closed-form sampling and is not used as a numerical discretization.\par

In the 3D flexible pipe case (Sec.~\ref{sec: cases: flow: 3D}), the flow domain is enclosed by the structure. We therefore sample directly from a tetrahedral mesh that evolves in time with nodal motion predicted from $\boldsymbol{\beta}$. As in 2D, element volumes determine sampling probabilities and barycentric sampling is used to generate spatial points consistent with the instantaneous geometry.\par

% Structure surface
\subsection{Structure Surface Sampling}
\label{sec: sampling: structure}
To evaluate the boundary loss, we sample points from the deforming interface $\partial \mathcal{V}_t$, discretized into line segments (2D) or triangles (3D). Segment endpoints and triangle vertices are updated at each time from the predicted surfacen points $\boldsymbol{p}$. Sampling along line segments is performed via uniform interpolation,
\begin{equation*}
    \boldsymbol{x} = (1 - r)\,\boldsymbol{x}_\mathrm{A} + r\,\boldsymbol{x}_\mathrm{B},
\end{equation*}
with $r \sim \mathcal{U}(0,1)$, and the associated surface velocity is also interpolated,
\begin{equation*}
    \boldsymbol{v}_{\boldsymbol{\beta}}(\boldsymbol{x}) =
    (1 - r)\,\boldsymbol{v}_{\boldsymbol{\beta}}(\boldsymbol{x}_\mathrm{A}) +
    r\,\boldsymbol{v}_{\boldsymbol{\beta}}(\boldsymbol{x}_\mathrm{B}).
\end{equation*}
For 3D surfaces, we apply barycentric sampling and interpolate the surface velocity using the same barycentric weights. In addition, for the 2D test case with a fixed circular boundary, we uniformly sample points along the circle to enforce the no-slip condition. These fixed-boundary samples remain constant in time.\par

% Particle tracks
\subsection{Particle Track Sampling}
\label{sec: sampling: particles}
The particle physics loss is evaluated along a batch of randomly selected tracks. Each track is defined over an interval $\sdx{\mathcal{T}}[k]$. For each selected track, we draw random times $t \in \sdx{\mathcal{T}}[k]$. Positions and velocities at these times are computed from $\sdx{\mathsf{P}}[k]$, and the corresponding fluid velocities are determined from $\mathsf{F}$. The loss is evaluated based on local discrepancies between particle and fluid velocities.\par

%%% Flow Cases %%%
\section{Flow Cases and Implementation Details}
\label{sec: cases}
We evaluate the reconstruction framework on three FSI problems: (1)~a 2D vortex-induced flow over a flexible beam (``flapping plate flow''), (2)~a 3D pulsatile flow through a compliant pipe (``flexi-pipe flow''), and (3)~a 3D wake generated by a fish (``swimming fish flow''). Ground truth data are generated from FSI simulations, and synthetic particle tracks are obtained by advecting Lagrangian tracers in the resulting velocity fields.\par

% Simulations
\subsection{FSI Simulations}
\label{sec: cases: flow}

% 2D flapping plate flow
\subsubsection{2D flapping plate Flow}
\label{sec: cases: flow: 2D}
The first test case is a classical benchmark for 2D FSI, featuring vortex-induced vibrations of a flexible plate (or beam) affixed to the rear of a cylindrical bluff body within a confined channel \cite{Turek2010}. This configuration exhibits strong two-way coupling and large structural deformations and is widely used to validate FSI solvers. The fluid domain is a channel measuring 2.5~m in length and 0.41~m in height. A rigid cylinder of radius 0.05~m is fixed at the vertical midpoint $(0.2~\text{m}, 0.2~\text{m})$, and a linearly elastic beam of length 0.35~m and thickness 0.02~m is attached downstream. Fluid enters from the left with a fully developed laminar velocity profile and mean speed 2~m/s, yielding $Re \approx 200$ based on the cylinder diameter. The inlet velocity is ramped up using a step function, and a Gaussian perturbation is introduced at $t = 1.5$~s to trigger vortex shedding. A stress-free condition is imposed at the outlet. The beam is modeled as a linear elastic solid with Young's modulus $E_\mathrm{s} = 5.6$~MPa, Poisson's ratio 0.4, and density $\rho_\mathrm{s} = 1000$~kg/m$^3$.\par

The simulation is conducted in COMSOL Multiphysics~5.6 using an ALE formulation. Fluid tractions along the beam induce structural motion, which feeds back to the fluid through a moving-wall condition. The computational mesh consists of 5856 elements, with unstructured triangles around the cylinder and plate to resolve boundary layers and wake dynamics; structured quadrilateral elements are used downstream of the plate. We use second-order Lagrange elements for velocity and pressure. Time integration uses a second-order backward differentiation formula with fixed time step $\Delta t = 0.05$~s. Stabilization is provided by a streamline upwind Petrov--Galerkin method, and the coupled system is solved using a Newton method with Anderson acceleration. The simulation runs for 5~s, with outputs recorded at 0.05~s intervals. We store the full velocity and pressure fields and the beam surface motion at each output time.\par

% 3D flexi-pipe flow
\subsubsection{3D Flexi-Pipe Flow}
\label{sec: cases: flow: 3D}
The second test case simulates internal pulsatile flow through a compliant, straight-walled cylindrical vessel to mimic large-vessel hemodynamics \cite{Liu2018}. The setup features two-way coupling between pressure-driven flow and wall deformation, resulting in pulse-wave propagation. The computational domain is a 10~cm long pipe with an inner radius of 1~cm (fluid lumen) and an outer radius of 1.2~cm (solid wall). The initial condition is quiescent. A step increase in pressure of 5~kPa is applied at the inlet at $t=0$~s to initiate flow. Stress-free boundary conditions are applied at the outlet and on the exterior wall. Axial displacements of the wall are constrained to zero, and a no-slip condition is enforced along the fluid--structure interface. The vessel wall is modeled as a homogeneous, isotropic, nearly incompressible neo-Hookean solid with density $\rho_\mathrm{s} = 1000$~kg/m$^3$, Young's modulus $E_\mathrm{s} = 1.0$~MPa, and Poisson's ratio 0.3. The fluid is assumed incompressible and Newtonian, with density $\rho = 1000$~kg/m$^3$ and dynamic viscosity $\mu = 0.004$~Pa~s.\par

Simulations are carried out in SimVascular \cite{Updegrove2017} using an ALE-based two-way coupled FSI solver. The lumen is discretized with 204{,}705 tetrahedral elements and the wall with 85{,}220 elements. The fluid and solid subproblems are solved sequentially at each time step. Structural dynamics are integrated using the generalized-$\alpha$ method, and the fluid equations use a second-order time discretization. The coupled nonlinear system is solved using GMRES with an incomplete LU preconditioner and threshold-based filtering. Mesh motion in the fluid domain is computed via harmonic extension of the wall displacement. The simulation is run for 0.02~s, and synthetic measurements are generated over $t \in [0, 0.012]$~s, during which the pressure wave propagates across most of the domain. Again, we store the full velocity and pressure fields and the wall deformation over the measurement interval.\par

% 3D swimming fish flow 
\subsubsection{3D Swimming Fish Flow}
\label{sec: cases: flow: fish}
The third and final test case considers a 3D wake flow driven by an undulatory swimmer. We use the high-fidelity dataset of Pan and Lauder~\cite{Pan2024}, which features a single ``giant danio'' fish swimming by carangiform locomotion. To reduce the geometric complexity of this scenario, the fish model retains the trunk and caudal fin while omitting the pelvic, dorsal, and pectoral fins. The fish has length $L = 0.07$~m and undulates according to a prescribed traveling-wave kinematics,
\begin{equation*}
    z(x, t) = A(x)\,\sin\!\left(\frac{2\pi}{\lambda}x - \frac{2\pi}{T}\,t\right),
\end{equation*}
with wavelength $\lambda = L$ and amplitude envelope $A(x) = 0.16\,x^2 - 0.08\,x + 0.02$ \cite{Menzer2025}. A uniform freestream velocity is imposed at the inlet, with a zero-gradient condition at the outlet and zero-stress conditions on the other boundaries. The fluid is incompressible and Newtonian, with density $\rho = 1000$~kg/m$^3$ and dynamic viscosity $\mu = 0.001$~Pa~s. Based on the freestream velocity and body length, the Reynolds number is $Re = 6396$ and the Strouhal number is $St = 0.386$.\par

Simulations are carried out with the solver of Menzer et al.~\cite{Menzer2025}, which solves the 3D unsteady incompressible Navier--Stokes equations on a nonconformal Cartesian grid using a cell-centered collocated arrangement. An immersed boundary method is used to represent the fish. Time advancement is done with a fractional-step scheme, using second-order Adams--Bashforth discretization for the convective terms and second-order Crank--Nicolson discretization for the diffusive terms. Near-body and wake regions are resolved using tree-topological local mesh refinement. We focus on the innermost block, which resolves the near-body wake of the fish. The corresponding subdomain spans $2.34L \times 0.64L \times 0.96L$ (approximately $0.164 \times 0.045 \times 0.067~\mathrm{m^3}$) and is discretized with $770 \times 226 \times 354$ cells in the streamwise ($x$), vertical ($y$), and spanwise ($z$) directions. For particle tracking, the velocity field is subsampled to $389 \times 57 \times 89$ to reduce computational cost while retaining the dominant wake structures. The simulation covers one full tail-beat cycle, sampled at 121 equally spaced snapshots.\par

% LPT track simulation
\subsection{Synthetic Particle Track Generation}
\label{sec: cases: tracks}
To simulate LPT measurements, we generate tracks that emulate key features of data processed using the Shake-The-Box pipeline \cite{Schanz2016}. Massless tracer particles of negligible size are uniformly seeded within the fluid domain and advected according to the simulated velocity field. Particle trajectories are computed using second-order Runge--Kutta integration with a time step consistent with the flow solver output. A periodic boundary condition is applied at the outlet, and particles that encounter solid boundaries are terminated and replaced by newly seeded tracers at the inlet.\footnote{Particle--structure collisions in these simulations arise from numerical integration error (e.g., when the advection step is not sufficiently small near boundary layers) or from under-resolved velocity gradients near solid surfaces. We treat these events as a practical source of track error, analogous to the loss of visibility or tracking failures in experiments.} Tracks with fewer than five time steps are discarded to reflect typical limits on detection and tracking.\par

\begin{figure}[htb!]
    \vspace*{-0.25em}
    \centering
    \includegraphics[width=0.9\textwidth]{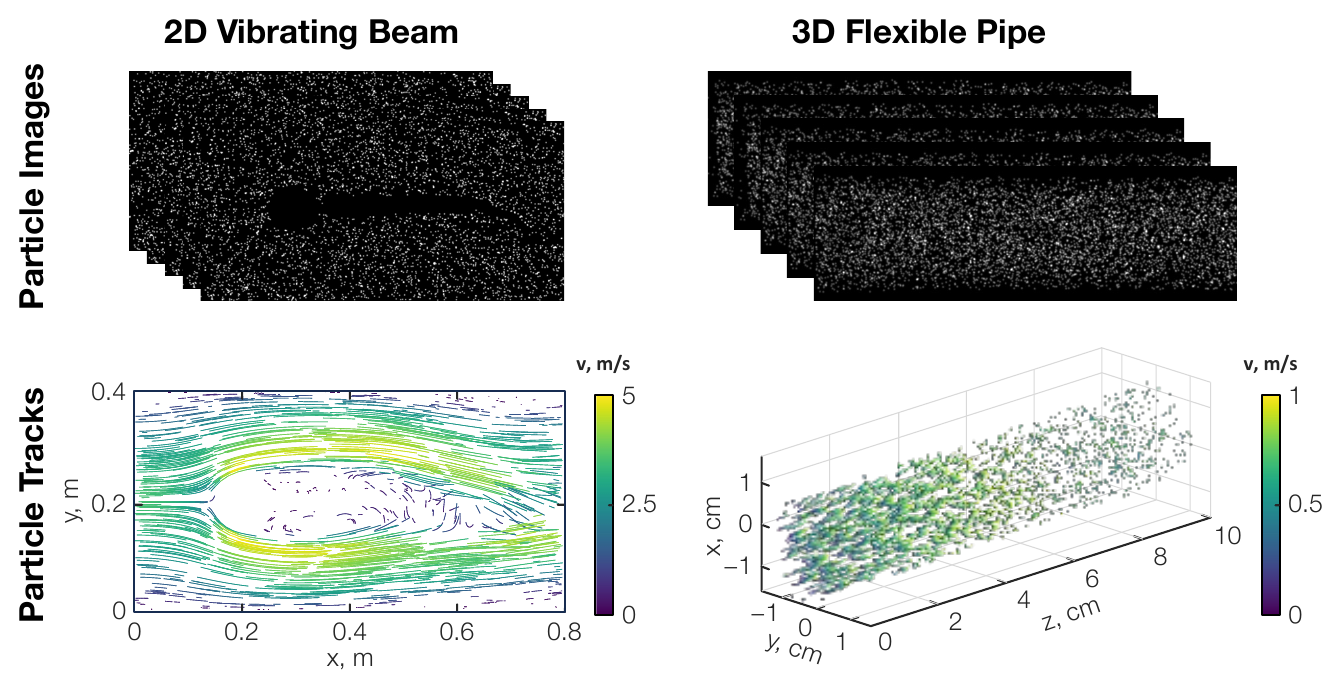}
    \caption{Synthetic particle tracks for the 2D flapping plate (left) and 3D flexi-pipe (right) cases. Particles are colored by instantaneous speed. Track density decreases near moving surfaces, limiting direct near-wall inference from raw tracks.}
    \label{fig: tracks}
\end{figure}

Figure~\ref{fig: tracks} presents representative synthetic images and reconstructed particle tracks for the 2D and 3D test cases. In the 2D flapping plate flow, particles are seeded at the inlet. We initialize 4000 tracers uniformly throughout $\mathcal{V}_0$ and advect them over 700 time steps with sampling interval $\Delta t = 0.05$~s. This corresponds to an effective particle image density of approximately 0.004~particles-per-pixel (ppp), assuming imaging with a 1~MP camera operating at 20~Hz. To maintain approximately steady sampling density, terminated tracks (e.g., due to wall contact) are replaced by newly seeded tracers. The same methodology is used for the 3D flexi-pipe flow. A total of 10,000 tracers are initialized uniformly throughout the lumen and advected for 120 frames. This configuration yields a particle density of approximately 0.01~ppp, assuming a multi-camera setup with 1~MP sensors operating at 10~kHz over 0.012~s.\footnote{This sampling rate is higher than in many laboratory systems; it is used here to resolve the short time window of pulse-wave propagation in this benchmark. The impact of the sampling rate is not explored in this study.} As in the 2D case, tracers that contact the vessel wall are removed and re-injected at the inlet. For flow around the swimming fish, we initialize $5 \times 10^4$ tracers uniformly within the subsampled fluid domain and advect them over one tail-beat cycle with dimensionless time step $\Delta t = 1/120$. This roughly corresponds to 0.05~ppp for a 1~MP camera sensor. This seeding density is comparable to values reported in volumetric PTV studies \cite{Schanz2016} and is sufficient to resolve the dominant wake structures in this dataset. Tracers that intersect the undulating fish body are removed and re-injected to maintain approximately steady sampling density.\par

We emphasize that the reconstructions in Sec.~\ref{sec: results} do not require each particle to be observed continuously over the full time horizon. Instead, long tracks are split into shorter segments to balance accuracy and computational cost. In our implementation, each trajectory is divided into fixed-length segments of 11 positions, with shorter segments zero-padded. Longer segments increase the dimension of coefficient matrices in the particle kinematics models and therefore increase memory use, whereas very short segments reduce temporal continuity and can introduce boundary-related artifacts. In practice, segment lengths of 8--15 points provide a reasonable balance between efficiency and fidelity.\par

%  Architecture and training
\subsection{Network Architecture and Training Protocol}
\label{sec: cases: numerics}
The governing flow field satisfies the incompressible Navier--Stokes equations, forming the physics residuals $\boldsymbol{e}_\mathrm{f}$ in the fluid domain $\mathcal{V}_t$. At the fluid--structure interface $\partial \mathcal{V}_t$, we impose a moving-wall boundary condition in weak form, yielding residuals $\boldsymbol{e}_\mathrm{s}$. For particle tracking, we assume purely advective motion, so the residuals for each particle track are $\sdx{\boldsymbol{e}}[k][\mathrm{p}] = \boldsymbol{u} - \sdx{\boldsymbol{v}}[k]$. The model includes four sets of trainable parameters: those of the velocity network $\mathsf{F}_{\boldsymbol{u}}$, pressure network $\mathsf{F}_p$, structure network $\mathsf{S}$, and particle kinematics models $\sdx{\mathsf{P}}[k]$. Network architectures vary across cases to accommodate differences in flow complexity. For the 2D flapping plate and 3D flexi-pipe cases, the velocity and pressure networks each use six hidden layers with 100 neurons per layer. The structure network uses four hidden layers with 150 neurons per layer. For the 3D swimming fish case, the velocity and pressure networks each use nine hidden layers with 150 neurons per layer, while the structure network again uses four hidden layers with 150 neurons per layer. To increase expressivity, we use Fourier encodings with 256 features. For the 2D case, spatial input frequencies are sampled from a zero-mean Gaussian distribution with standard deviation 0.2, and we use a standard deviation of 0.8 for time. A standard deviation of 0.2 is used for all frequencies in the pipe case, and for the fish case, spatial and temporal frequencies are sampled with standard deviations of 2 and 0.2, respectively. Network weights are initialized from a standard normal distribution, with biases set to zero.\par

Loss weights are chosen through a coarse grid search and kept constant during training. All the weights are set to unity except for the particle physics loss, which couples the track data to the flow field and varies by case. We set this weight to $\gamma_3 = 10$ for the flapping plate problem, $\gamma_3 = 10^5$ for the 3D flexi-pipe case, and $\gamma_3 = 10^5$ for the fish case. We also impose a soft penalty on volumetric expansion for the fish trunk, equivalent to $\mathrm{d}|\mathcal{V}_t|/\mathrm{d}t = 0$, to discourage non-physical deformations. Since biological tissue is nearly incompressible, the enclosed body volume is expected to remain approximately constant over the swimming cycle. The volume loss weight is also set to unity.\par

For each training iteration in the 2D case, 2000 samples are drawn from the fluid domain $\Omega$: 1800 from the fixed subregions and 200 from the dynamic near-field mesh, sampled across 25 randomly chosen time frames. For the surface loss, 200 boundary points are sampled from $\partial \mathcal{V}_t$ per frame over 25 frames, yielding 5000 points per iteration. For the particle loss, we first select 5000 tracks with probabilities proportional to track length. From each selected track, 10 random times $t \in \sdx{\mathcal{T}}[k]$ are drawn, and the corresponding positions are computed using $\sdx{\mathsf{P}}[k]$ via Eq.~\eqref{equ: continuous particle trajectory}. In the flexi-pipe case, we use 5000 points from the fluid volume, 5000 from the structure boundary, and 5000 samples drawn from tracks, each evaluated at 10 random times. For the fish flow reconstruction, the training batch consists of 5000 fluid collocation points sampled across 10 randomly selected frames, with 450 points drawn from the fixed region and 50 from the near-body zone per frame. We also sample 5000 points on the fish surface, consisting of 400 points per frame on the fish trunk and 100 points per frame on the caudal fin, again over 10 frames. Additionally, the particle loss is evaluated using 5000 samples from the trajectory ensemble. A discussion of these sampling choices is provided in Sec.~\ref{sec: results: sampling}. All parameters are updated using the Adam optimizer with a fixed learning rate of $10^{-3}$. Training is conducted until convergence, requiring approximately $2 \times 10^5$ iterations for the flapping plate and swimming fish cases and $10^5$ iterations for the flexi-pipe case. The framework is implemented in TensorFlow~2.10 and executed on an NVIDIA GeForce RTX~3090 GPU. The total training time is approximately 5.5~hours for the 2D case, 1.5~hours for the 3D flexi-pipe case, and 6.5~hours for the 3D swimming fish case.\par

%%% Results %%%
\section{Results}
\label{sec: results}
Here, we evaluate the FSI reconstruction framework on the benchmarks described above. Reconstructions are performed using the particle tracks and a prescribed modal basis for the structures. We first summarize how to obtain these bases in Sec.~\ref{sec: results: modes}. We then present FSI reconstructions for each case in Secs.~\ref{sec: results: plate}--\ref{sec: results: fish}. Robustness to localization errors is examined in Sec.~\ref{sec: results: noise}. Finally, we analyze the sampling scheme used to approximate the volume and surface integrals in Sec.~\ref{sec: results: sampling}.\par

% Modes
\subsection{Modal Representations}
\label{sec: results: modes}
\begin{figure}[b!]
    \centering    
    \includegraphics[width=0.9\textwidth]{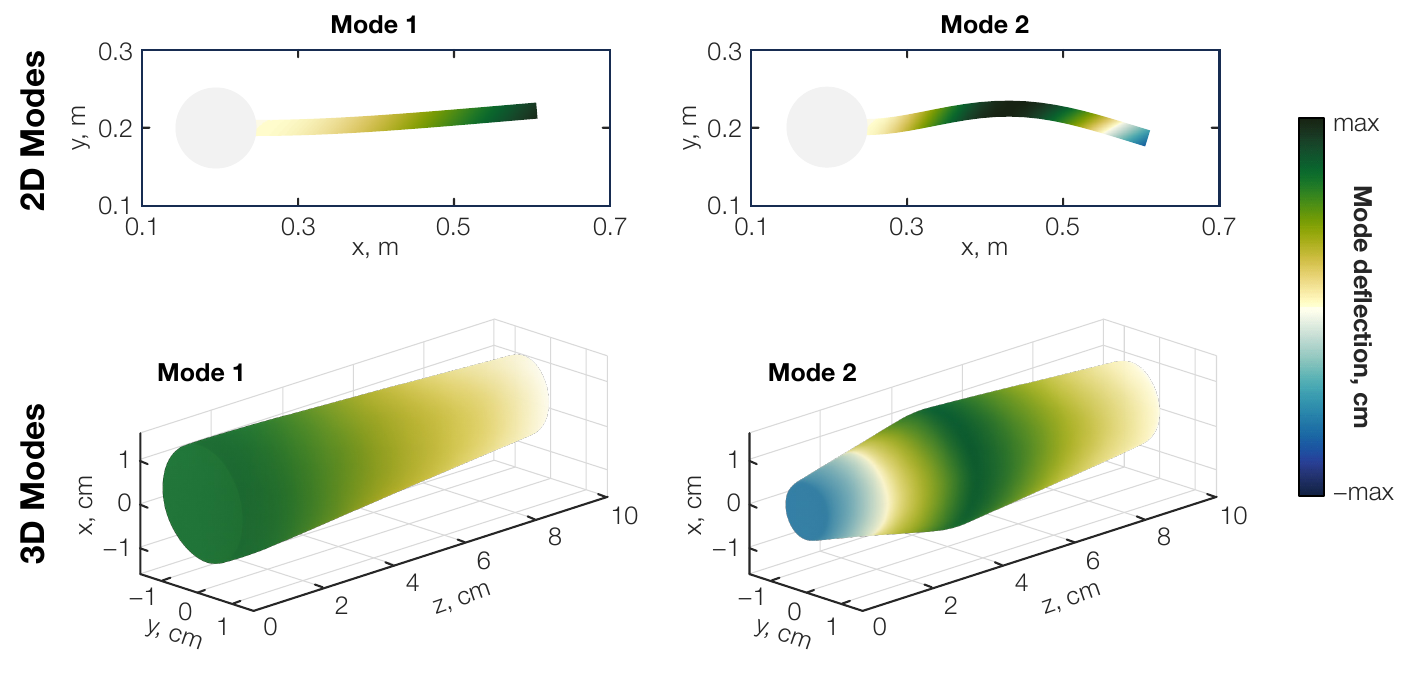}
    \caption{Representative deformation modes for the 2D flapping plate (top) and 3D flexible pipe (bottom).}
    \label{fig: modes}
\end{figure}

A modal representations is employed for the deformable surface $\partial \mathcal{V}_t$. A surface mesh is represented by a vector of nodal positions $\boldsymbol{p}$, decomposed into a base structure $\overline{\boldsymbol{p}}$ and a time-dependent deformation field $\boldsymbol{p}^\prime(t) = \boldsymbol{p}(t) - \overline{\boldsymbol{p}}$, as described in Sec.~\ref{sec: method: model: structure}. Surface deformations are approximated in a low-dimensional subspace, $\boldsymbol{p}^\prime(t) \approx \boldsymbol{\Phi}\boldsymbol{\beta}(t)$, where the columns of $\boldsymbol{\Phi}$ are spatial deformation modes and $\boldsymbol{\beta}$ is the vector of modal coefficients outputted by $\mathsf{S}$. We consider two strategies for constructing the basis. One is a physics-based eigenmode analysis, which we demonstrate for the 2D case. The other is a data-driven POD basis, which we employ for both 3D cases. Leading modes for the 2D plate and 3D pipe are shown in Fig.~\ref{fig: modes}.\par

Deformation modes for the flapping plate are computed through an eigenfrequency analysis of the cylinder--beam assembly in COMSOL. The structure is modeled as a linear, isotropic elastic solid, with the beam clamped to the rear of a rigid cylinder. The free-vibration problem,
\begin{equation*}
    \boldsymbol{M} \frac{\mathrm{d}^2 \boldsymbol{p}^\prime}{\mathrm{d} t^2} + \boldsymbol{K} \boldsymbol{p}^\prime = \boldsymbol{0},
\end{equation*}
is cast as the generalized eigenvalue problem
\begin{equation*}
    \left(\boldsymbol{K} - \omega_j^2 \boldsymbol{M}\right)\boldsymbol{\varphi}_j = \boldsymbol{0},
\end{equation*}
where $\boldsymbol{M}$ and $\boldsymbol{K}$ are the global mass and stiffness matrices, $\omega_j$ is the natural circular frequency of the $j$th mode, and $\boldsymbol{\varphi}_j$ is the associated eigenvector. Although these eigenmodes are not guaranteed to coincide with the most energetic deformation patterns in the coupled FSI response (as POD modes would do), they are physically interpretable and easy to compute. In our 2D reconstructions, we retain the first eight modes. The leading two modes capture approximately 99.85\% of the deformation energy (as measured by the variance of $\boldsymbol{p}^\prime$), and eight modes capture 99.98\% of the energy. The top row of Fig.~\ref{fig: modes} shows the first two bending modes.\par

For the flexible pipe and fish cases, we represent the surface response using a data-driven POD basis, consistent with the continuous formulation in Sec.~\ref{sec: method: model: structure}. Deformation snapshots from the simulation are assembled into a matrix of surface displacements, $\boldsymbol{P}^\prime = [\boldsymbol{p}^\prime(t_0), \boldsymbol{p}^\prime(t_1), \dots]$, and we apply a singular value decomposition,
\begin{equation*}
    \boldsymbol{P}^\prime = \boldsymbol{\Phi}\boldsymbol{\Sigma}\boldsymbol{\Psi}^\top.
\end{equation*}
Here, $\boldsymbol{\Phi}$ contains spatial deformation modes ordered by energy. POD bases are compatible with experimental workflows (e.g., using DIC), since they can be constructed from sparse or asynchronous surface measurements. In the pipe case, we retain between 1 and 14 modes, accounting for 44.5\% and 99\% of the deformation energy, respectively; the first two modes are shown in the bottom row of Fig.~\ref{fig: modes}. For the swimming fish, we retain between 1 and 6 modes, capturing from 63\% to 99.9\% of the energy.\par

% flapping plate flow
\subsection{2D Flapping Plate Flow Reconstructions}
\label{sec: results: plate}
We first evaluate our framework on the canonical flapping plate benchmark. Structural deformations are represented using up to eight bending modes, as described above. Particle tracks are generated as outlined in Sec.~\ref{sec: cases: tracks} and shown on the left side of Fig.~\ref{fig: tracks}. The seeding density is lowest near the plate due to the cylinder wake. Moreover, steep gradients near the plate can reduce the accuracy of numerical particle advection in this region. Together, these effects produce spatially varying errors that are most pronounced about the interface. Hence, motion of the plate must be inferred from off-body observations through joint enforcement of the flow physics and the moving-wall condition.\par

\begin{figure}[b!]
    \vspace*{-0.25em}
    \centering
    \includegraphics[width=0.9\textwidth]{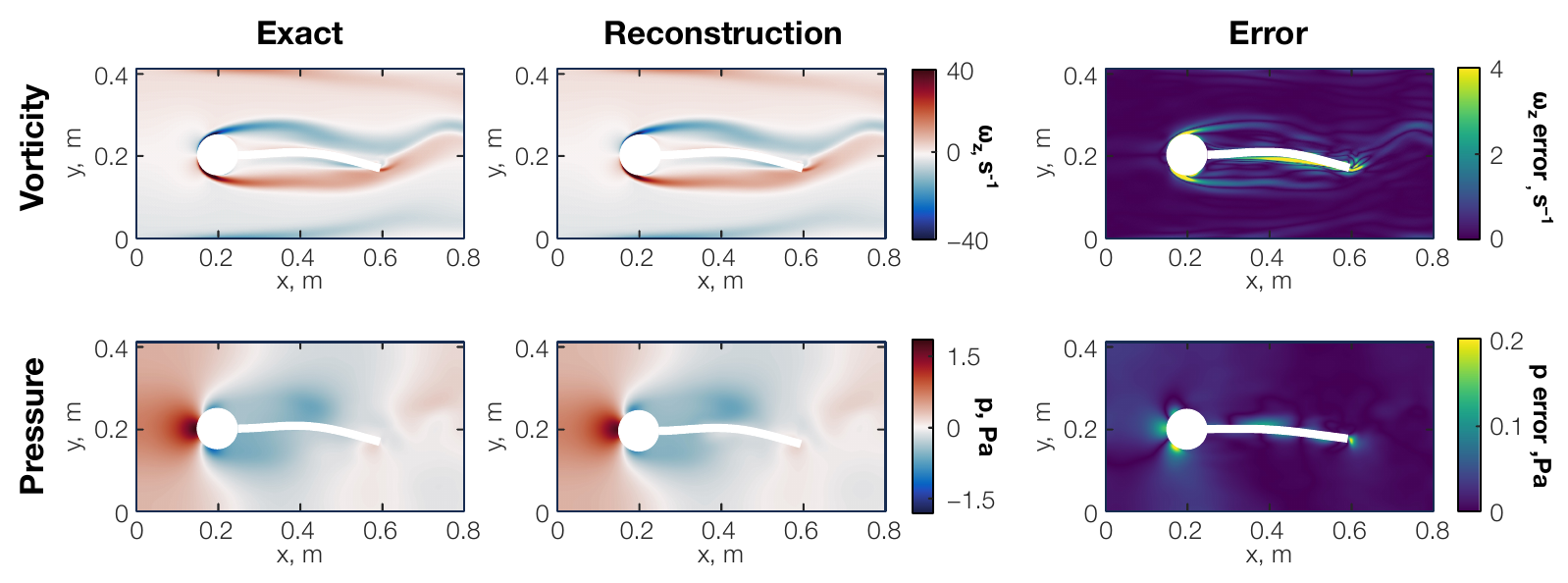}
    \caption{Flapping plate case: exact (left), reconstructed (middle), and error (right) fields for vorticity (top) and pressure (bottom).}
    \label{fig: 2D reconstructions}
\end{figure}

Figure~\ref{fig: 2D reconstructions} shows a representative snapshot of the ground truth and reconstructed vorticity and pressure fields. For consistency, both fields are projected onto a structured mesh using a reference PINN trained directly on the ground truth data. This step is used because the CFD solution is produced on a moving, unstructured mesh that conforms to the plate, while the reconstructed plate may differ slightly from the ground truth, so the native solution points do not necessarily align with the reconstruction grid. The reconstruction captures the main features of the flow well, including vortex shedding and low-pressure regions in the cylinder wake. Normalized root-mean-square errors (NRMSEs) for vorticity and pressure are approximately 15\% and 8\%, respectively. Errors are elevated along the shear layer and near the surface, consistent with regions of acute data sparsity and stronger gradients. Across the modal configurations tested, reconstruction accuracy remains comparable, suggesting that inferred flow fields are only weakly sensitive to the inclusion of additional (nearly redundant) structural modes.\par

\begin{figure}[htb!]
    \centering
    \includegraphics[width=0.9\textwidth]{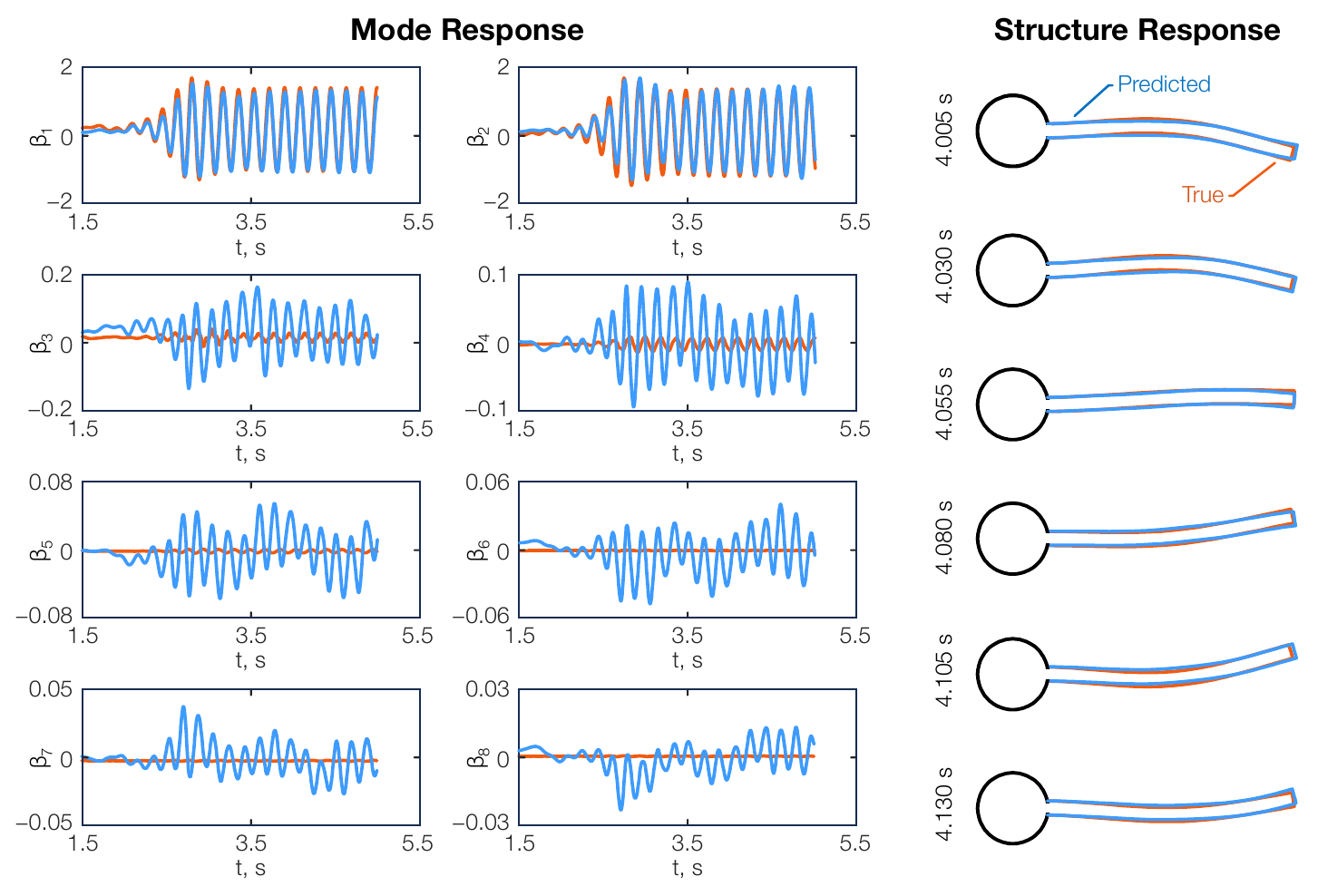}
    \caption{Flapping plate structure: exact (red) and inferred (blue) mode coefficients (left) and reconstructed surface shapes at six time instants (right).}
    \label{fig: 2D structure}
\end{figure}

Figure~\ref{fig: 2D structure} depicts the time-varying mode coefficients inferred by the structure network for the eight-mode case. The leading mode pair, which accounts for more than 99.85\% of the deformation energy, is recovered with quite good agreement in amplitude and phase. Higher-order modes exhibit larger deviations, consistent with their smaller energetic contribution. The reconstructed surface shapes nonetheless closely track the true motion of the plate.\par

\begin{figure}[htb!]
    \vspace*{-0.25em}
    \centering
    \includegraphics[width=0.85\textwidth]{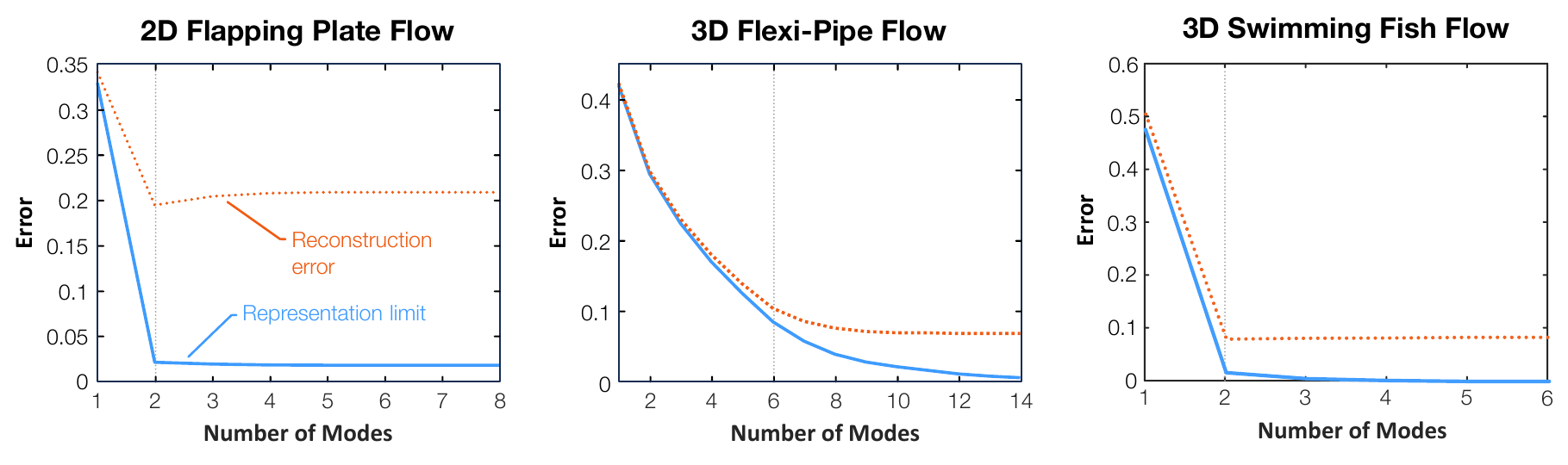}
    \caption{Representation and reconstruction errors versus modal truncation for the 2D flapping plate (left), 3D flexi-pipe (middle), and 3D swimming fish (right) cases. Errors plateau after the dominant deformation energy is captured (vertical dashed line), indicating limited sensitivity to additional modes.}
    \label{fig: mode sensitivity}
\end{figure}

Sensitivity to modal truncation is summarized in the left-most panel of Fig.~\ref{fig: mode sensitivity}. Each data point corresponds to an independent reconstruction using the number of modes indicated on the $x$-axis. We compare reconstruction error to the representation error,
\begin{equation*}
    \frac{1}{n_\mathrm{t}} \sum_{j=1}^{n_\mathrm{t}}
    \frac{\left\lVert \left(\boldsymbol{I} - \boldsymbol{\Pi}\right) \boldsymbol{p}^\prime(t_j) \right\rVert_{\boldsymbol{M}}^2}
         {\left\lVert \boldsymbol{p}^\prime(t_j) \right\rVert_{\boldsymbol{M}}^2},
\end{equation*}
where $\boldsymbol{\Pi}$ denotes the mass-weighted projection onto the span of the retained modes,
\begin{equation*}
    \boldsymbol{\Pi} =
    \boldsymbol{\Phi}\left(\boldsymbol{\Phi}^\top \boldsymbol{M}\boldsymbol{\Phi}\right)^{-1}\boldsymbol{\Phi}^\top \boldsymbol{M}.
\end{equation*}
For POD bases computed with the Euclidean inner product, we set $\boldsymbol{M} = \boldsymbol{I}$, in which case $\boldsymbol{\Pi} = \boldsymbol{\Phi}\boldsymbol{\Phi}^\top$. The representation error measures the fraction of deformation energy not captured by the selected modal subspace; it is thus a lower bound on reconstruction error for this norm. In the flapping plate case, the representation error drops below 2\% after two modes, indicating that the deformation time series is effectively low dimensional. Correspondingly, reconstruction error decreases when the second mode is added and then varies weakly with additional modes. Adding modes beyond this point provides limited additional benefit; at the same time, it does not meaningfully degrade performance in this case. This behavior is helpful in experimental settings, where the required number of modes may not be known a priori. Hence, robustness to over-specification is desirable.\par

% Pipe flow
\subsection{3D Flexi-Pipe Flow Reconstructions}
\label{sec: results: pipe}
We next test the framework on a 3D scenario, featuring pulsatile flow through a compliant vessel. Structural deformations are represented using a POD basis extracted from the simulation. As in the 2D case, particle tracks are sparsely distributed throughout the domain and are particularly scarce near the pipe walls.\par

\begin{figure}[b!]
    \centering
    \includegraphics[width=.9\textwidth]{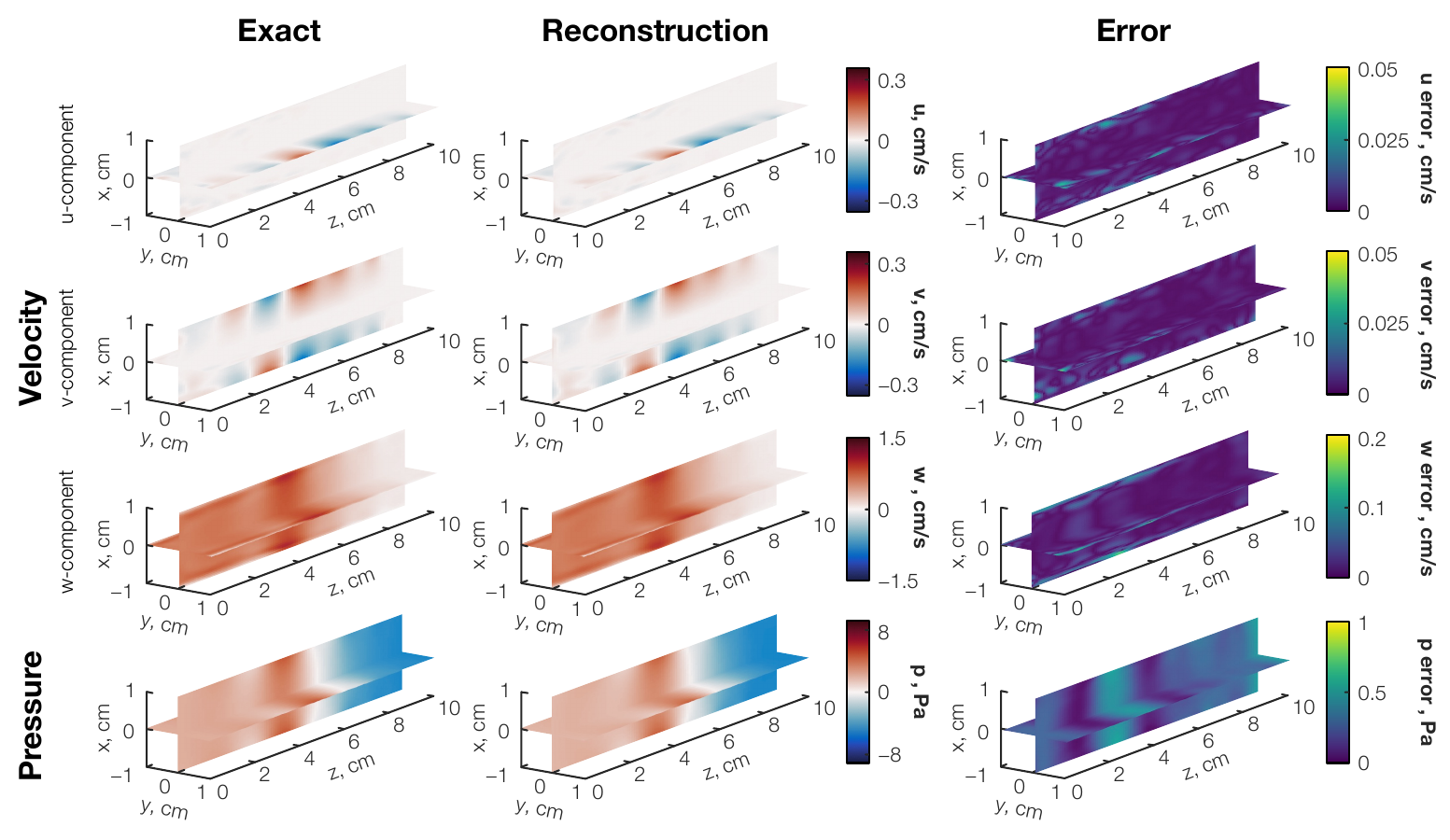}
    \caption{Flexi-pipe case: exact (left), reconstructed (middle), and error (right) fields for velocity (top three rows) and pressure (bottom) on a representative slice.}
    \label{fig: 3D reconstructions}
\end{figure}

Figure~\ref{fig: 3D reconstructions} presents reconstructed flow fields at a representative snapshot. The reconstruction captures the main flow features, including transverse vortical structures near the vessel wall and axial propagation of a pressure-driven wave along the lumen. Time-averaged NRMSEs are approximately 14\% for the transverse velocity components, 5\% for axial velocity, and 9\% for pressure. The largest discrepancies occur near the wall, where sparse particle coverage and stronger gradients reduce the effective information content of the LPT observations. Across modal truncations beyond six modes, the flow reconstruction error varies weakly, holding roughly constant around 9\% in this case.\par

\begin{figure}[tb!]
    \centering
    \includegraphics[width=.9\textwidth]{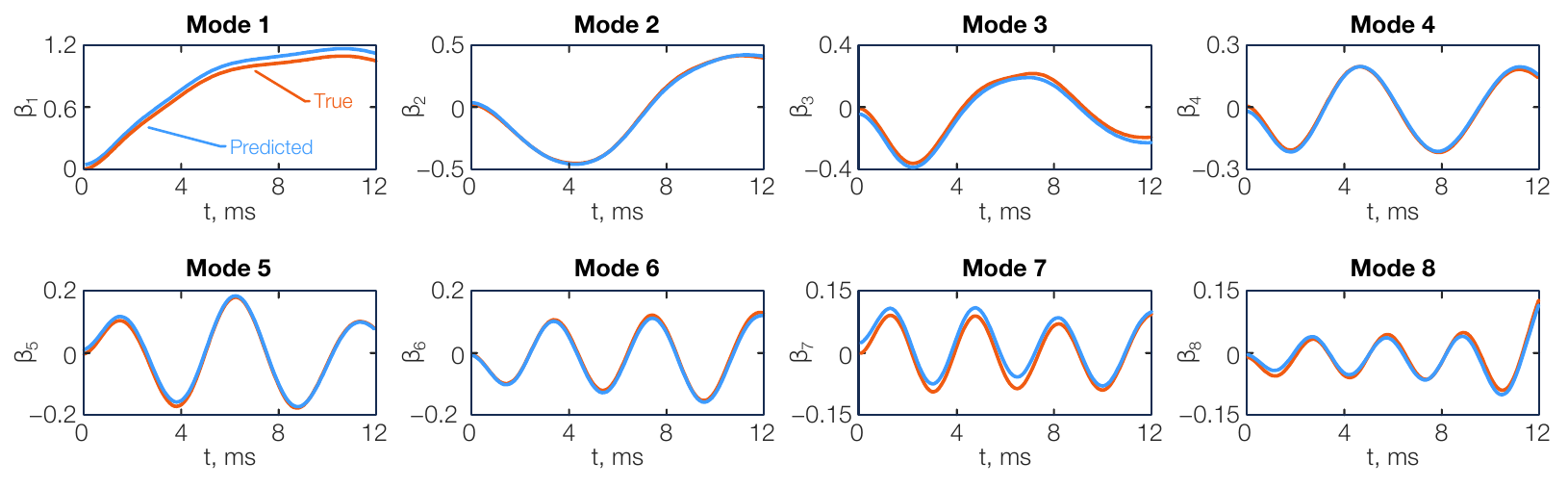}
    \caption{Flexi-pipe case: inferred (blue) and exact (red) POD mode coefficients versus time.}
    \label{fig: 3D coefficients}
\end{figure}

\begin{figure}[b!]
    \vspace*{-0.25em}
    \centering
    \includegraphics[width=0.9\textwidth]{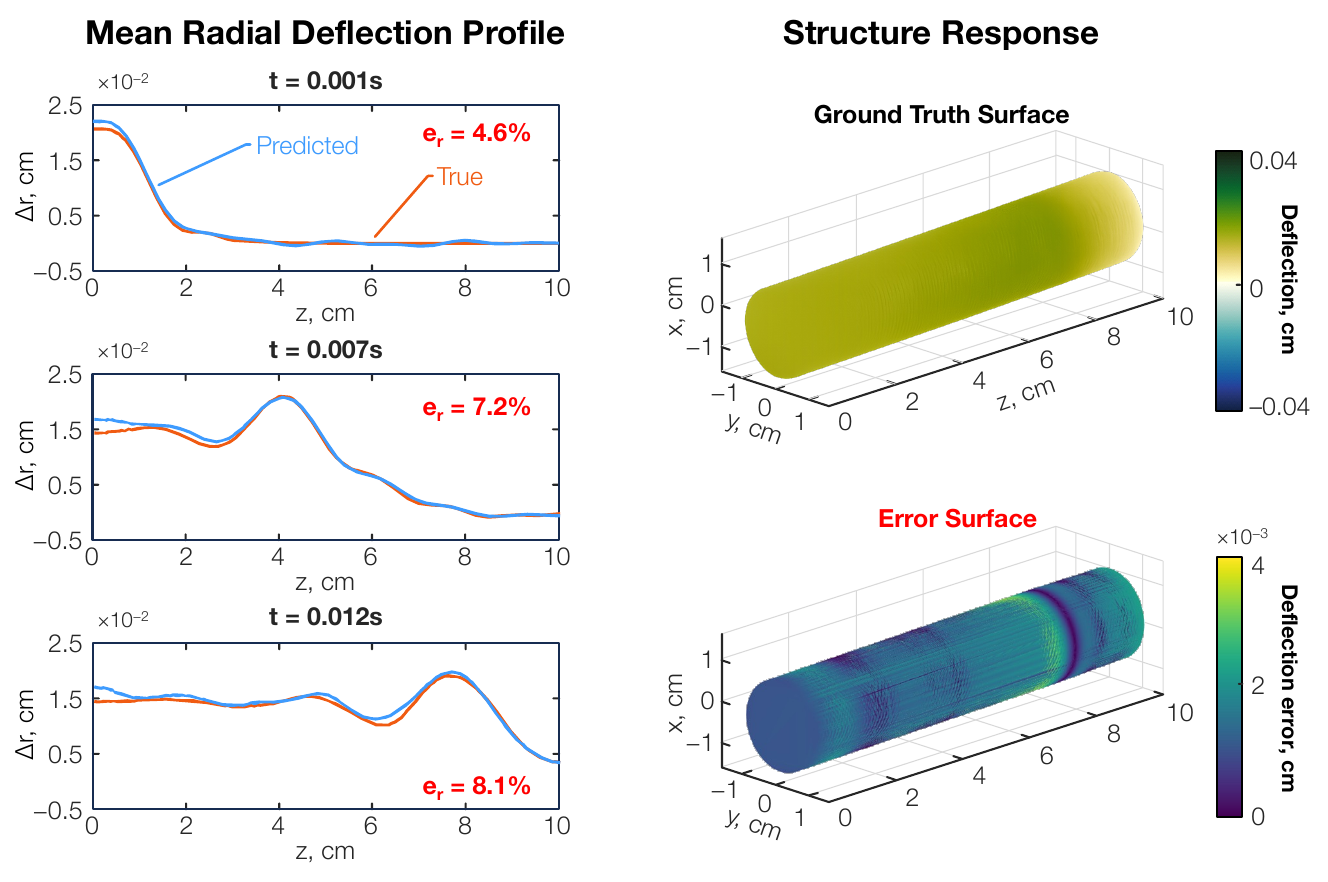}
    \caption{Flexi-pipe surface reconstruction. Left: radial deflection profiles at three time instants. Right: ground-truth deformation and corresponding error field at the final frame.}
    \label{fig: 3D structure}
\end{figure}

Figure~\ref{fig: 3D coefficients} compares the inferred POD mode coefficients to the truth values. Good agreement is achieved for all eight modes, including both periodic and aperiodic time series. Figure~\ref{fig: 3D structure} shows the corresponding surface reconstruction. The left panel reports radial deflection profiles at three representative time instants, computed as circumferential averages of the radial displacement at each axial location. The predicted profiles track the ground truth at these times, with relative errors of approximately 4.6\%, 7.2\%, and 8.1\%. Errors increase modestly over time as the wave propagates and gradients sharpen, consistent with the limited near-wall sampling. Overall, the surface deformation is recovered with relative errors remaining below 9\%. The right panel shows the surface deformation at the final time, colored by radial deflection magnitude and overlaid with the corresponding error field; the predicted deflection reproduces the characteristic bulge traveling along the pipe, consistent with the reconstructed pressure wave in Fig.~\ref{fig: 3D reconstructions}.\par

The middle panel of Fig.~\ref{fig: mode sensitivity} summarizes sensitivity to modal truncation for the pipe flow case. Structural reconstruction errors decrease rapidly with the number of retained POD modes and plateau beyond approximately eight modes, which collectively capture over 95\% of the deformation energy. Reconstruction accuracy closely tracks the representation limit of the basis up to about six modes, after which performance varies weakly with additional modes. Hence, for this case, the method is again (relatively) insensitive to over-specification.\par

% 3D swimming fish flow
\subsection{3D Swimming Fish Flow Reconstructions}
\label{sec: results: fish}
The final flow case features the flow around a swimming fish, introducing a more complex surface geometry and chaotic wake dynamics. Once again, the surface motion is represented using a POD basis extracted from the simulation. The first two modes account for 94\% of the deformation energy and primarily represent the traveling-wave kinematics imposed on the fish. Higher-order modes capture finer scale motions of the fins. As in the previous cases, particle tracks are sparse near the moving surface. Reconstruction is further complicated by the nominally zero-thickness caudal fin, which produces steep near-surface gradients that are difficult to constrain from off-body observations.\par

\begin{figure}[htb!]
    \centering
    \includegraphics[width=.9\textwidth]{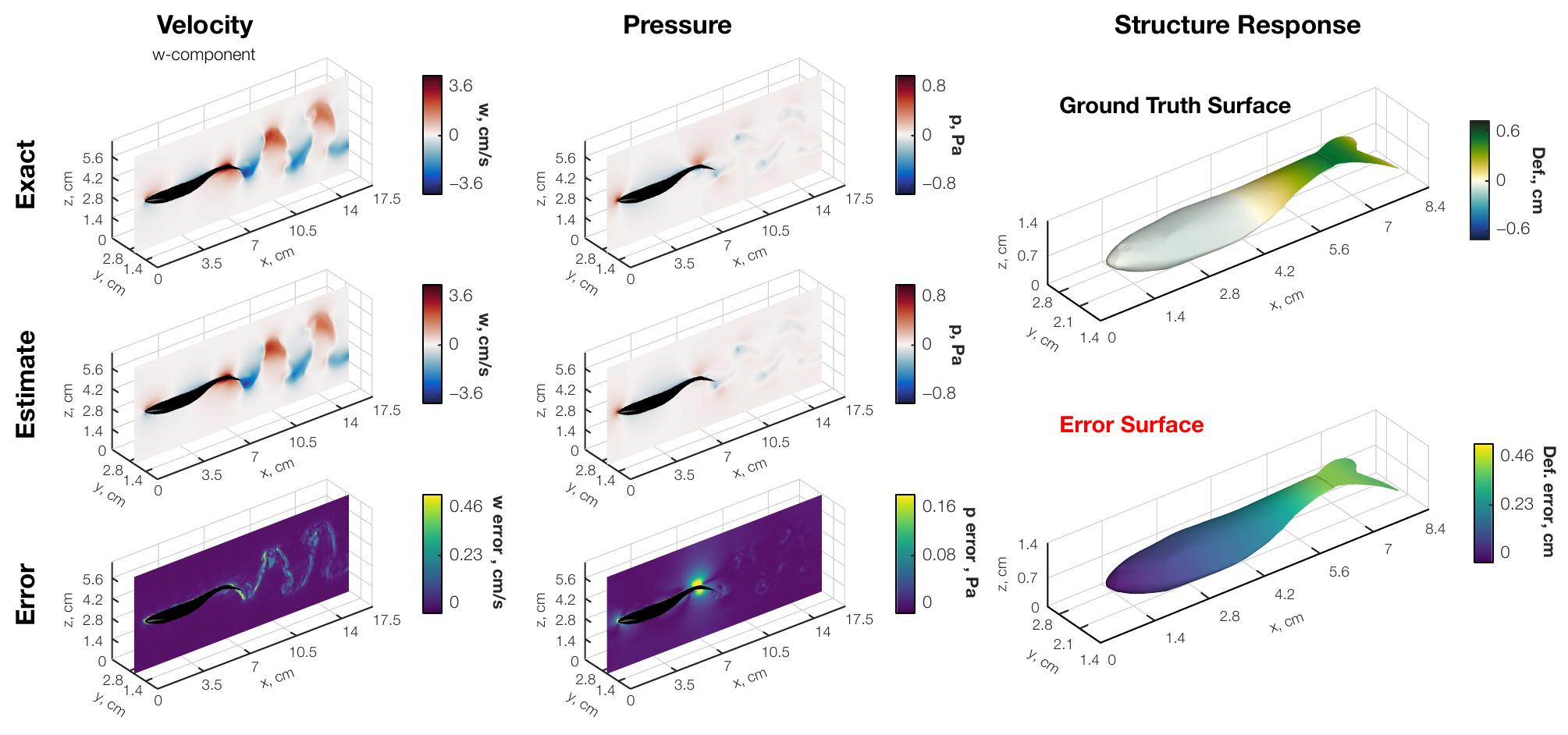}
    \caption{Swimming fish case: exact (top), reconstructed (middle), and error (bottom) fields for spanwise velocity ($w$) (left) and pressure (center), with ground-truth surface deformation and associated error (right).}
    \label{fig: fish reconstructions}
\end{figure}

Figure~\ref{fig: fish reconstructions} compares the reconstructed flow and structure fields at a representative instant during the tail-beat cycle. The reconstruction recovers the main wake features, as can be seen in the spanwise velocity cut plots, including coherent vortical structures shed from the caudal fin. The pressure field is qualitatively consistent with these structures, exhibiting low-pressure regions that coincide with the vortex cores. Quantitatively, the time-averaged relative error is approximately 13\% for the transverse velocity components, 0.6\% for the streamwise component, and 17\% for pressure. As in the plate and pipe cases, the errors peak near the moving boundary and in the near-field region of the wake, wherein gradients are strongest and the density of particles is lowest. Discrepancies are particularly concentrated near the caudal fin. This behavior is consistent with the sharp velocity gradients induced by the thin fin geometry and the tendency of coordinate neural representations to favor smooth solutions, which can limit accuracy in the vicinity of such features.\par

Sensitivity to the number of modes, shown in the right-most panel of Fig.~\ref{fig: mode sensitivity}, exhibits a similar trend to the other cases. Reconstruction error for the fish body decreases rapidly as modes are added, dropping from approximately 50\% with one mode to below 10\% after two. Beyond this point, errors vary weakly (approximately 8\%) as additional modes are incorporated. Across the tested truncations, we do not observe large-amplitude spurious structure motion with the inclusion of higher-order modes, suggesting, as before, that the reconstruction is relatively insensitive to over-specification.\par

% Noisy datasets
\subsection{Robustness to Particle Localization Error}
\label{sec: results: noise}
Localization error is a common source of uncertainty in LPT experiments \cite{Schroder2023}. We assess sensitivity to such errors by generating noisy datasets for each case at increasing levels of noisy. We then compare (i)~raw track velocities obtained by finite differences and (ii)~jointly optimized tracks produced by the proposed framework, and we quantify the effects of noise on the accuracy of the inferred tracks, flow fields, and structural motion.\par

% Noisy generation
\subsubsection{Generating Noisy Data}
\label{sec: results: noise: model}
Estimated particle positions in LPT are subject to uncertainty due to calibration error, centroid detection limits, and optical distortions \cite{Schanz2016}. We model these effects by perturbing the synthetic particle positions with additive, zero-mean Gaussian noise. Let $\sdx{\boldsymbol{x}}[k][i]$ denote the true particle position at time $t_i$ for track $k$. We generate noisy observations as
\begin{equation}
    \label{equ: noise model}
    \sdx{\widetilde{\boldsymbol{x}}}[k][i] = \sdx{\boldsymbol{x}}[k][i] + \sdx{\boldsymbol{\epsilon}}[k][i],
\end{equation}
where $\sdx{\boldsymbol{\epsilon}}[k][i]$ is drawn from a centered Gaussian distribution with covariance $\boldsymbol{\Gamma}=\mathrm{diag}(\sigma_1^2,\ldots,\sigma_d^2)$, which encodes anisotropic localization uncertainty. Localization errors are often larger along the least well-resolved measurement direction (e.g., the out-of-plane coordinate in multi-camera reconstructions).\par

For the 2D case, we apply low, mid, and high noise levels with $(\sigma_x,\sigma_y)=(0.1, 0.2)$, $(0.5, 1)$, and $(1, 2)$ pixels. For the flexi-pipe case, we set $(\sigma_{yz}, \sigma_x) = (0.1, 0.2)$, $(0.5, 1)$, and $(1, 2)$ pixels. Lastly, for the fish case, we set $(\sigma_{xy}, \sigma_z) = (0.1, 0.2)$, $(0.5, 1)$, and $(1, 2)$ pixels. These noise levels span a range intended to reflect practical LPT conditions, based on sub-pixel accuracies commonly reported for multi-camera STB systems \cite{Schanz2016} to larger uncertainties encountered in single-camera volumetric techniques such as digital in-line holography \cite{Mallery2019, Shao2020}. The prescribed anisotropy ratio of 2 between the well-resolved (in-plane) and poorly-resolved (out-of-plane) directions is consistent with reported trends for volumetric reconstructions \cite{Shao2020, Moaven2024}.\par

% Tracks
\subsubsection{Track Refinement Results}
\label{sec: results: noise: tracks}

\begin{figure}[b!]
    \centering
    \includegraphics[width=.9\textwidth]{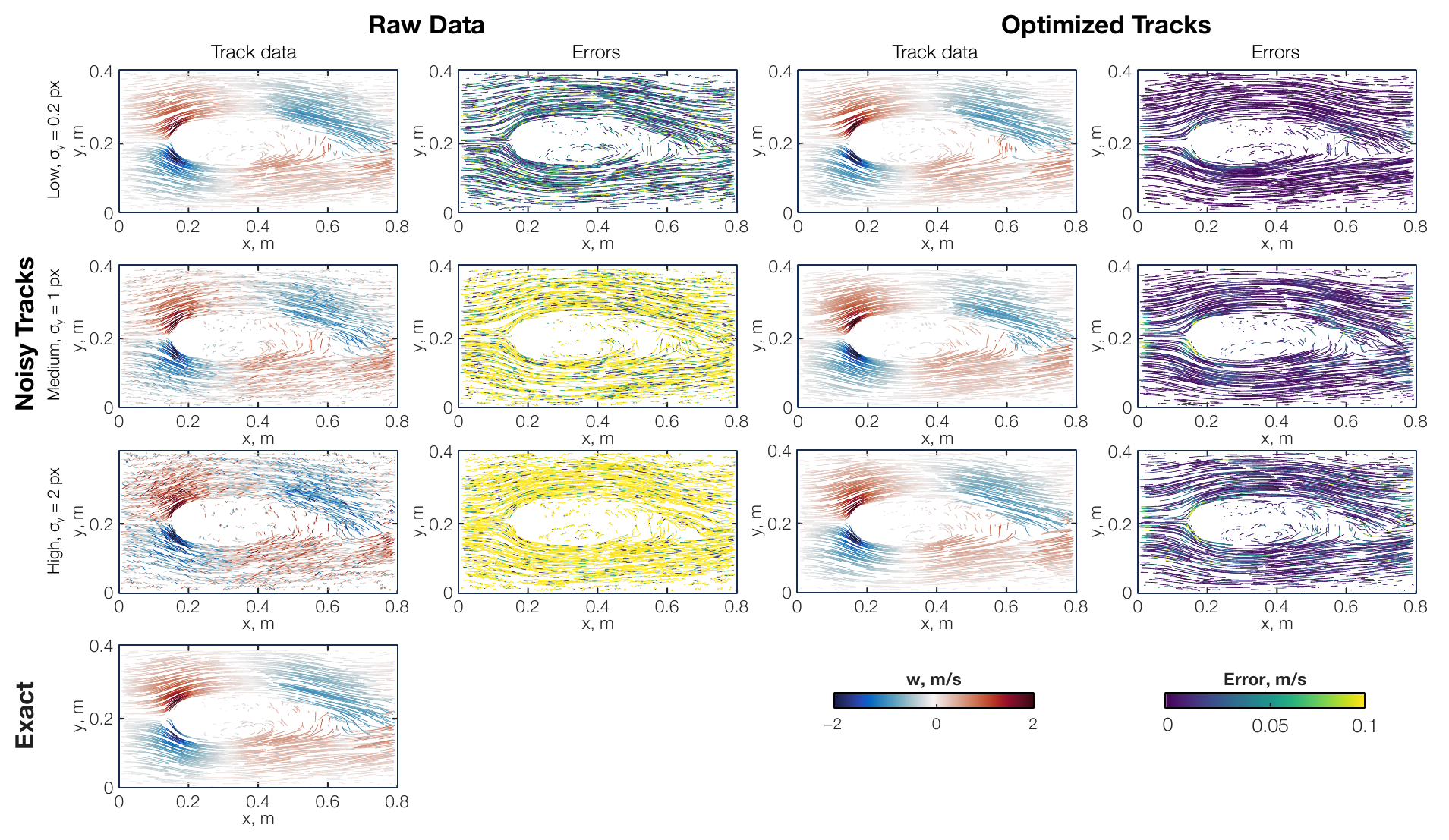}
    \caption{Effect of localization error on track quality for the flapping plate case. Rows: $\sigma_y = 0.2$, $1$, and $2$ pixels (top to bottom). Left: raw tracks and error fields. Right: optimized tracks and error fields. Tracks are colored by the vertical velocity component ($v$). The bottom row shows the noise-free reference.}
    \label{fig: tracks panel}
\end{figure}

When particles are subject to finite localization uncertainty, we estimate their positions (subject to the assumed uncertainty model) alongside the flow field and surface motion. We refer to this here as \emph{trajectory refinement}, which can be viewed as a form of physics-informed particle tracking. For each noise level, we rerun the reconstruction using the same procedure as in the noise-free cases, except that the covariance $\boldsymbol{\Gamma}$ in the data loss is set to be consistent with the prescribed values of $\sigma_x$, $\sigma_y$, and $\sigma_z$,\footnote{In principle, one could also infer the $\sigma_i$ within a Bayesian formulation; we leave this for future work.} as opposed to setting $\mathscr{J}_\mathrm{data} = 0$. Figure~\ref{fig: tracks panel} shows this behavior for the flapping plate case. As the level of noise increases, the raw track data yield increasingly erratic velocities. After optimization, the tracks become smoother and more consistent with the inferred flow, with the largest reductions in error occurring near the moving boundary, i.e., where raw track errors are greatest. This reflects the combined roles of the data term, which discourages overfitting to noisy positions, and the coupled physics losses, which constrain the trajectories via consistency with the flow field.\par

Table~\ref{tab: noise} shows reconstruction errors for track velocities, flow fields, and structural motion across all three benchmarks as a function of localization noise. Raw track errors, computed by finite differencing the noisy positions, grow with the level of noise. The flexi-pipe case yields particularly large raw errors because its characteristic velocity scale is small, making finite-difference estimates sensitive to localization uncertainty. In all cases, following joint optimization of the track, flow field, and structure models, the track errors decrease substantially.\par

\begin{table}[htb!]
    \renewcommand{\arraystretch}{1.15}
    \setlength{\tabcolsep}{6pt}
    \caption{Reconstruction errors subject to localization uncertainties. Track velocity errors are normalized by a characteristic velocity scale for each case. ``Raw'' errors are computed by applying finite differences to noisy position data; ``Optimized'' errors are obtained after the joint reconstruction. Flow field and structural errors are reported for the maximum number of modes.}
    \centering
    \begin{tabular}{c c c c c c c}
        \hline\hline
        \multirow{2}{*}{\bf Case} & \multirow{2}{*}{\bf Noise level} & \multicolumn{2}{c}{\bf Track velocity} & \multicolumn{2}{c}{\bf Flow field} & \multirow{2}{*}{\bf Structure} \\
        & & Raw & Optimized & $\|\boldsymbol{v}\|$ & $p$ & \\[.1em]
        \hline
        \multirow{4}{*}{2D flapping plate}
        & Clean & --    & --    & 0.029 & 0.081 & 0.20 \\
        & Low   & 0.025 & 0.009 & 0.029 & 0.098 & 0.21 \\
        & Mid   & 0.100 & 0.012 & 0.030 & 0.108 & 0.22 \\
        & High  & 0.200 & 0.019 & 0.031 & 0.103 & 0.22 \\[4pt]
        \multirow{4}{*}{3D flexi-pipe}
        & Clean & --    & --    & 0.042 & 0.093 & 0.07 \\
        & Low   & 0.320 & 0.012 & 0.049 & 0.096 & 0.07 \\
        & Mid   & 1.920 & 0.032 & 0.064 & 0.107 & 0.15 \\
        & High  & 4.050 & 0.063 & 0.067 & 0.103 & 0.17 \\[4pt]
        \multirow{4}{*}{3D swimming fish}
        & Clean & --    & --    & 0.006 & 0.172 & 0.07 \\
        & Low   & 0.042 & 0.003 & 0.007 & 0.238 & 0.08 \\
        & Mid   & 0.208 & 0.019 & 0.007 & 0.222 & 0.08 \\
        & High  & 0.416 & 0.019 & 0.008 & 0.259 & 0.08 \\
        \hline\hline
    \end{tabular}
    \label{tab: noise}
\end{table}

% Flow
\subsubsection{Flow Results}
\label{sec: results: noise: flow}
Table~\ref{tab: noise} also reports flow reconstruction errors caused by localization error for a representative modal configuration (i.e., with the maximum number of retained modes). Results for other mode counts are similar, consistent with the truncation insensitivity observed in the noise-free study. Across all three benchmarks, velocity errors remain bounded as noise increases. The 2D plate case shows only a modest change in velocity error (from $2.9\%$ for clean data to $3.1\%$ highly noisy data). The 3D flexi-pipe case shows the largest increase (from $4.2\%$ to $6.7\%$), and the swimming fish case remains below $0.8\%$ across noise levels.\par

Pressure errors are higher than velocity errors across all cases, which is emblematic of PINN-based reconstructions of flow fields. Even so, pressure error growth remains moderate across all our tests. For the 2D plate, pressure error increases from $8.1\%$ to $10.8\%$; for the 3D flexi-pipe case, it remains within $9.3\%$--$10.7\%$; and for the fish, pressure errors are higher overall and range from $17.2\%$ to $25.9\%$, due to acute errors about the tail fin.\par

% Structure
\subsubsection{Structural Response}
\label{sec: results: noise: structure}
The 2D plate and 3D fish cases show limited sensitivity to particle localization noise, with displacement errors increasing by less than 2\% from the noise-free baseline to the highest level. The pipe case is more sensitive, increasing from 7\% to 17\%. This behavior is consistent with the smaller deformation amplitude in this pressure-driven flow: radial wall displacements are only a few percent of the pipe radius, yielding a lower signal-to-noise ratio for structural inference. By comparison, the swimming fish and flapping plate undergo larger relative deformations that remain more readily observable under the imposed noise. Despite this increased sensitivity, the reconstructed wall motion in the flexi-pipe case still captures downstream propagation of the pressure response.\par

% Sampling scheme
\subsection{Analysis of the Sampling Scheme}
\label{sec: results: sampling}
All volume and surface loss terms are approximated using Monte Carlo integration. The accuracy of these estimates, and the stability of training, depend on the sampling strategy: variance in the sampled loss propagates to gradient estimates during backpropagation. High variance can induce oscillatory behavior near optima or prevent convergence. Low-variance estimates of the continuous loss terms thus promote stable training.\par

\begin{figure}[htb!]
    \vspace*{-0.25em}
    \centering
    \includegraphics[width=0.9\textwidth]{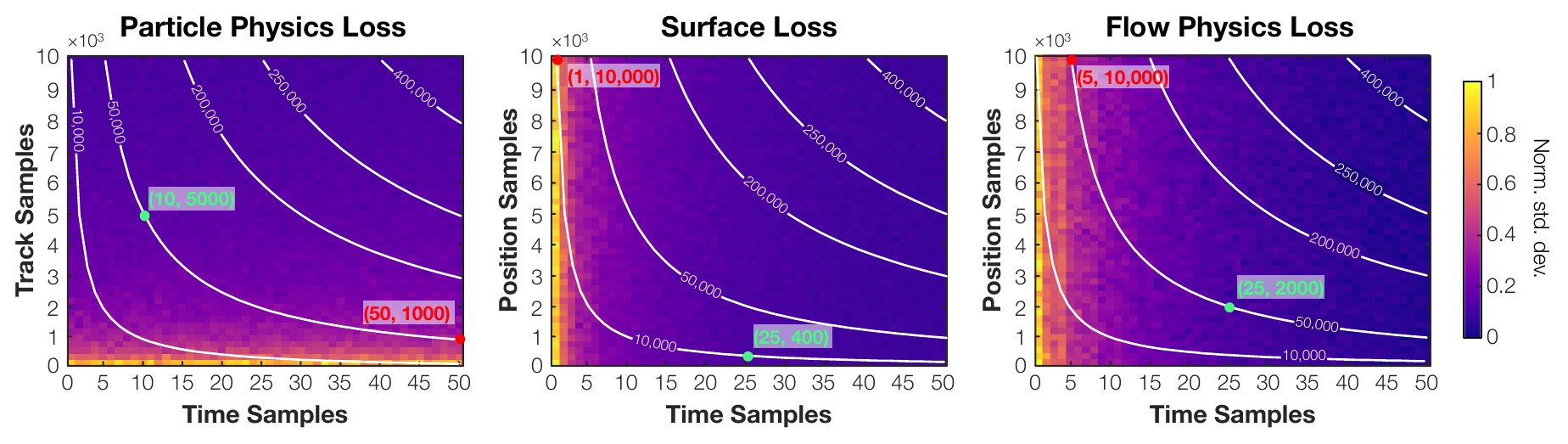}
    \vspace*{0.5em}
    \caption{Normalized standard deviation maps for the particle (left), boundary (center), and flow physics (right) loss terms in the 2D flapping plate case. Axes denote the number of tracks, spatial samples, or temporal samples per iteration. White curves indicate fixed sample budgets. Green and red dots mark representative low- and high-variance sampling configurations.}
    \label{fig: loss variance}
\end{figure}

To characterize sampling-induced variance, we evaluate a range of batch configurations using a randomly initialized network with fixed weights and biases. This provides a conservative baseline for variance and isolates the effects of sampling from changes in the model parameters during training. For a grid of spatial and temporal sample sizes, we compute 100 independent Monte Carlo estimates of the flow physics loss, surface boundary loss, and particle physics loss for the 2D flapping plate case. For each configuration, we report the standard deviation of each loss term, normalized by its maximum over the grid, yielding the variance maps in Fig.~\ref{fig: loss variance}. These results guide batch-size selection by highlighting trade-offs between a reduction in variance and the computational cost of integration.\par

Trends in Fig.~\ref{fig: loss variance} show that increasing the number of time samples per iteration reduces variance in the boundary and flow physics losses, whereas the number of spatial samples has a weaker effect for this case. The particle physics loss is governed primarily by the number of distinct tracks which are sampled, with a weaker dependence on the number of samples drawn per track. White curves superimposed on the maps indicate fixed sample budget (a proxy for computational cost) and illustrate that the variance in a loss term can differ substantially while holding cost constant.\par

To connect the trends in variance with training outcomes, we reconstruct the 2D flapping plate case using two representative configurations with similar sample counts: a low-variance configuration (``good sampling'') and a high-variance configuration (``poor sampling''). These are marked by the green and red points in Fig.~\ref{fig: loss variance}. In the good case, we sample 5000 tracks and 10 time instants for $\mathscr{J}_\mathrm{part}$, 400 spatial positions and 25 times for $\mathscr{J}_\mathrm{surf}$, and 2000 positions and 25 times for $\mathscr{J}_\mathrm{flow}$. In the poor case, we sample 1000 tracks and 50 times for $\mathscr{J}_\mathrm{part}$, 10,000 positions and 1 time for $\mathscr{J}_\mathrm{surf}$, and 10,000 positions and 5 times for $\mathscr{J}_\mathrm{flow}$.\par

\begin{figure}[htb!]
    \vspace*{-0.25em}
    \centering
    \includegraphics[width=0.45\textwidth]{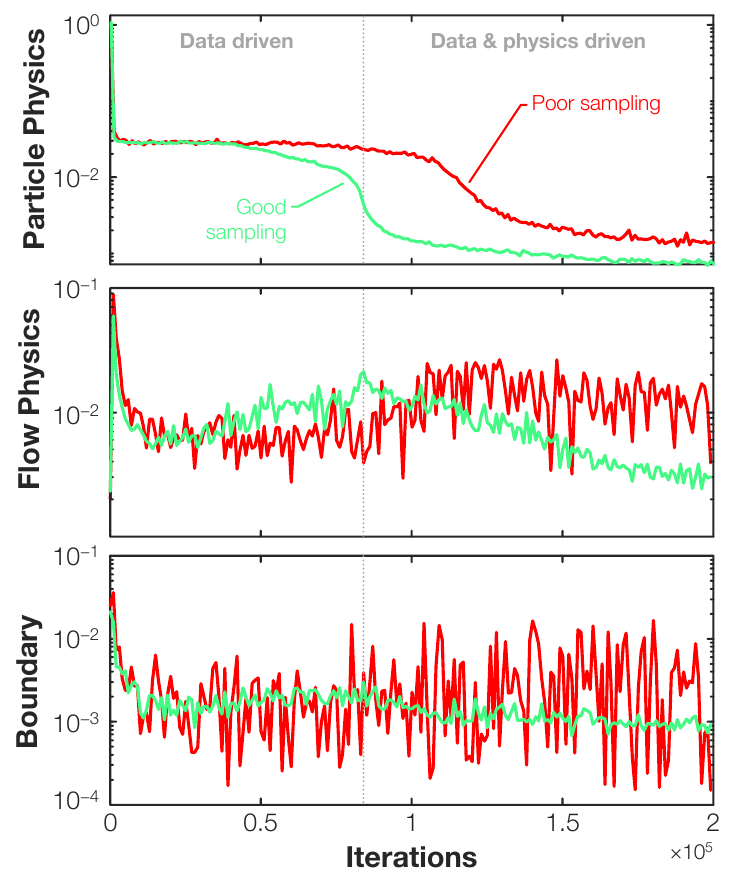}
    \caption{Evolution of component-wise loss terms during training using the low-variance (green) and high-variance (red) sampling strategies: particle physics loss (top), flow physics loss (middle), and boundary loss (bottom).}
    \label{fig: training loss}
\end{figure}

Evolution of the loss terms over training differs markedly between these strategies. Using the low-variance configuration, all loss components decrease smoothly with relatively small fluctuations. Training proceeds in stages: with an initial rapid decrease in all terms, followed by an interval in which the particle loss continues to decrease while the physics losses change more slowly. After approximately $8.5 \times 10^4$ iterations (marked by the vertical gray line in Fig.~\ref{fig: training loss}), the flow and boundary losses begin to decrease more consistently, indicating stronger coupling between the data and physics constraints. Similar staged behavior was reported by Molnar and Grauer~\cite{Molnar2022}. In contrast, the high-variance configuration leads to unstable training, despite the similar computational cost. The loss components exhibit large fluctuations, and the flow and boundary losses do not converge over the training window. After roughly $1.2 \times 10^5$ iterations, training enters a limit cycle in which short-lived improvements are followed by a rebound in error. Over the final 1000 iterations, the variances of the flow and boundary losses are 7.5-fold and 24-fold higher, respectively, than under the low-variance strategy. These training dynamics translate into reconstruction quality: the high-variance strategy yields final errors of 5.4\% in velocity, 17\% in pressure, and 50\% in the surface reconstruction, whereas the low-variance strategy reduces these errors to 3.5\%, 10\%, and 20\%, respectively, underscoring the importance of the sampling scheme.\par

%%% Conclusions %%%
\section{Conclusions}
\label{sec: conclusions}
This manuscript presents a physics-informed framework for reconstructing coupled flow and structural dynamics from sparse, single-phase, potentially noisy particle tracking data. The approach combines neural-implicit representations of the fluid and structure with a moving-wall condition, enabling joint inference of time-resolved flow fields and structural motions from off-body particle tracks. In the formulation studied here, the structure is represented through a prescribed deformation basis, and the method does not require direct structural measurements or explicit force data. This complements existing methodologies for reconstructing FSI, which typically rely on multi-modal measurements, dense volumetric flow data, or better-specified structural models.\par

We evaluated the framework on three synthetic benchmarks with LPT-style observations: vortex-induced oscillations of a 2D flapping plate, pressure-driven motion in a 3D compliant pipe, and a 3D wake driven by an undulatory swimming fish. In each case, reconstructions were performed using particle tracks and a prescribed modal basis, and the results were assessed against the simulated ground truth. Across all the tested conditions, the method recovered the dominant flow features and the main deformation dynamics well, with the largest discrepancies occurring near moving boundaries and in regions of acute data sparsity.\par

We also examined two practical aspects that influence performance. First, a modal truncation study showed that reconstruction accuracy improves as the dominant deformation modes are included and then varies weakly with additional modes, indicating a limited sensitivity to over-specification of the modal basis in these benchmarks. This is expected to be beneficial in experimental studies, where the optimal basis is not known a priori. Second, a noise study with perturbed particle positions demonstrated that the framework can accommodate moderate to high levels of localization error by jointly refining the particle trajectories and flow\slash structure states; performance degradaded in proportion to the signal-to-noise ratio of the deformation (most notably in the flexi-pipe case). Finally, an analysis of the sampling scheme highlighted how sampling-induced variance can affect training stability, underscoring the importance of allocating spatial and temporal batches in a manner consistent with the transient domains.\par

%%% Acknowledgments %%%
\section*{Acknowledgments}
This material is based upon work supported by the National Science Foundation under Grant No. 2501442.\par

%%% References %%%


\begin{thebibliography}{10}

\bibitem{Nakata2012}
T.~Nakata and H.~Liu, \enquote{{A fluid--structure interaction model of insect flight with flexible wings},} J. Comput. Phys. \textbf{231}, 1822--1847 (2012).

\bibitem{Zhu2021}
H.~Zhu, Q.~Sun, X.~Liu, J.~Liu, H.~Sun, W.~Wu, P.~Tan, and Z.~Chen, \enquote{{Fluid--structure interaction-based aerodynamic modeling for flight dynamics simulation of parafoil system},} Nonlinear Dyn. \textbf{104}, 3445--3466 (2021).

\bibitem{Calderer2018}
A.~Calderer, X.~Guo, L.~Shen, and F.~Sotiropoulos, \enquote{{Fluid--structure interaction simulation of floating structures interacting with complex, large-scale ocean waves and atmospheric turbulence with application to floating offshore wind turbines},} J. Comput. Phys. \textbf{355}, 144--175 (2018).

\bibitem{Korobenko2013}
A.~Korobenko, M.-C. Hsu, I.~Akkerman, J.~Tippmann, and Y.~Bazilevs, \enquote{{Structural mechanics modeling and FSI simulation of wind turbines},} Math. Models Methods Appl. Sci. \textbf{23}, 249--272 (2013).

\bibitem{Suito2014}
H.~Suito, K.~Takizawa, V.~Q. Huynh, D.~Sze, and T.~Ueda, \enquote{{FSI analysis of the blood flow and geometrical characteristics in the thoracic aorta},} Comput. Mech. \textbf{54}, 1035--1045 (2014).

\bibitem{Lee2017}
A.~H. Lee, R.~L. Campbell, B.~A. Craven, and S.~A. Hambric, \enquote{{Fluid--structure interaction simulation of vortex-induced vibration of a flexible hydrofoil},} J. Vib. Acoust. \textbf{139}, 041001 (2017).

\bibitem{Donea1982}
J.~Donea, S.~Giuliani, and J.-P. Halleux, \enquote{{An arbitrary Lagrangian-Eulerian finite element method for transient dynamic fluid-structure interactions},} Comput. Methods Appl. Mech. Eng. \textbf{33}, 689--723 (1982).

\bibitem{Hughes1981}
T.~J. Hughes, W.~K. Liu, and T.~K. Zimmermann, \enquote{{Lagrangian-Eulerian finite element formulation for incompressible viscous flows},} Comput. Methods Appl. Mech. Eng. \textbf{29}, 329--349 (1981).

\bibitem{Legay2006}
A.~Legay, J.~Chessa, and T.~Belytschko, \enquote{{An Eulerian--Lagrangian method for fluid--structure interaction based on level sets},} Comput. Methods Appl. Mech. Eng. \textbf{195}, 2070--2087 (2006).

\bibitem{Jenkins2015}
N.~Jenkins and K.~Maute, \enquote{{Level set topology optimization of stationary fluid-structure interaction problems},} Struct. Multidiscip. Optim. \textbf{52}, 179--195 (2015).

\bibitem{Yu2013}
Y.~Yu, H.~Baek, and G.~E. Karniadakis, \enquote{{Generalized fictitious methods for fluid--structure interactions: analysis and simulations},} J. Comput. Phys. \textbf{245}, 317--346 (2013).

\bibitem{Pathak2016}
A.~Pathak and M.~Raessi, \enquote{{A 3D, fully Eulerian, VOF-based solver to study the interaction between two fluids and moving rigid bodies using the fictitious domain method},} J. Comput. Phys. \textbf{311}, 87--113 (2016).

\bibitem{Peskin2002}
C.~S. Peskin, \enquote{{The immersed boundary method},} Acta Numer. \textbf{11}, 479--517 (2002).

\bibitem{Sotiropoulos2014}
F.~Sotiropoulos and X.~Yang, \enquote{{Immersed boundary methods for simulating fluid--structure interaction},} Prog. Aerosp. Sci. \textbf{65}, 1--21 (2014).

\bibitem{Hou2012}
G.~Hou, J.~Wang, and A.~Layton, \enquote{{Numerical methods for fluid-structure interaction---a review},} Commun. Comput. Phys. \textbf{12}, 337--377 (2012).

\bibitem{Haeri2012}
S.~Haeri and J.~Shrimpton, \enquote{{On the application of immersed boundary, fictitious domain and body-conformal mesh methods to many particle multiphase flows},} Int. J. Multiphase Flow \textbf{40}, 38--55 (2012).

\bibitem{Chatellier2013}
L.~Chatellier, S.~Jarny, F.~Gibouin, and L.~David, \enquote{{A parametric PIV/DIC method for the measurement of free surface flows},} Exp. Fluids \textbf{54}, 1--15 (2013).

\bibitem{Bleischwitz2017}
R.~Bleischwitz, R.~De~Kat, and B.~Ganapathisubramani, \enquote{{On the fluid-structure interaction of flexible membrane wings for MAVs in and out of ground-effect},} J. Fluids Struct. \textbf{70}, 214--234 (2017).

\bibitem{Zhang2019}
P.~Zhang, S.~D. Peterson, and M.~Porfiri, \enquote{{Combined particle image velocimetry/digital image correlation for load estimation},} Experimental Thermal and Fluid Science \textbf{100}, 207--221 (2019).

\bibitem{Hortensius2017}
R.~Hortensius, J.~C. Dutton, and G.~S. Elliott, \enquote{{Simultaneous planar PIV and sDIC measurements of an axisymmetric jet flowing across a compliant surface},} in \enquote{{55th AIAA Aerospace Sciences Meeting},}  (2017), p. 1886.

\bibitem{DAguanno2023}
A.~D'Aguanno, P.~Quesada~Allerhand, F.~F.~J. Schrijer, and B.~W. van Oudheusden, \enquote{{Characterization of shock-induced panel flutter with simultaneous use of DIC and PIV},} Exp. Fluids \textbf{64}, 15 (2023).

\bibitem{Kosters2023}
W.~I. K{\"o}sters and S.~Hoerner, \enquote{{Simultaneous flow measurement and deformation tracking for passive flow control experiments involving fluid--structure interactions},} J. Fluids Struct. \textbf{121}, 103956 (2023).

\bibitem{Mitrotta2022}
F.~M. Mitrotta, J.~Sodja, and A.~Sciacchitano, \enquote{{On the combined flow and structural measurements via robotic volumetric PTV},} Meas. Sci. Technol. \textbf{33}, 045201 (2022).

\bibitem{Schroder2023}
A.~Schr{\"o}der and D.~Schanz, \enquote{{3D Lagrangian particle tracking in fluid mechanics},} Annu. Rev. Fluid Mech. \textbf{55}, 511--540 (2023).

\bibitem{Elkins2007}
C.~J. Elkins and M.~T. Alley, \enquote{{Magnetic resonance velocimetry: applications of magnetic resonance imaging in the measurement of fluid motion},} Exp. Fluids \textbf{43}, 823--858 (2007).

\bibitem{Kontogiannis2022}
A.~Kontogiannis, S.~V. Elgersma, A.~J. Sederman, and M.~P. Juniper, \enquote{{Joint reconstruction and segmentation of noisy velocity images as an inverse Navier--Stokes problem},} J. Fluid Mech. \textbf{944}, A40 (2022).

\bibitem{Karnakov2024}
P.~Karnakov, S.~Litvinov, and P.~Koumoutsakos, \enquote{{Solving inverse problems in physics by optimizing a discrete loss: Fast and accurate learning without neural networks},} PNAS nexus \textbf{3}, pgae005 (2024).

\bibitem{Buhendwa2024}
A.~B. Buhendwa, D.~A. Bezgin, P.~Karnakov, N.~A. Adams, and P.~Koumoutsakos, \enquote{{Shape inference in three-dimensional steady state supersonic flows using ODIL and JAX-Fluids},} arXiv preprint arXiv:2408.10094  (2024).

\bibitem{Raissi2019a}
M.~Raissi, P.~Perdikaris, and G.~E. Karniadakis, \enquote{{Physics-informed neural networks: A deep learning framework for solving forward and inverse problems involving nonlinear partial differential equations},} J. Comput. Phys. \textbf{378}, 686--707 (2019).

\bibitem{Raissi2019b}
M.~Raissi, Z.~Wang, M.~S. Triantafyllou, and G.~E. Karniadakis, \enquote{{Deep learning of vortex-induced vibrations},} J. Fluid Mech. \textbf{861}, 119--137 (2019).

\bibitem{Kharazmi2021}
E.~Kharazmi, D.~Fan, Z.~Wang, and M.~S. Triantafyllou, \enquote{{Inferring vortex induced vibrations of flexible cylinders using physics-informed neural networks},} J. Fluids Struct. \textbf{107}, 103367 (2021).

\bibitem{Tang2022}
H.~Tang, Y.~Liao, H.~Yang, and L.~Xie, \enquote{{A transfer learning-physics informed neural network (TL-PINN) for vortex-induced vibration},} Ocean Eng. \textbf{266}, 113101 (2022).

\bibitem{Zhu2024}
Y.~Zhu, W.~Kong, J.~Deng, and X.~Bian, \enquote{{Physics-informed neural networks for incompressible flows with moving boundaries},} Phys. Fluids \textbf{36} (2024).

\bibitem{Sundar2024}
R.~Sundar, D.~Majumdar, D.~Lucor, and S.~Sarkar, \enquote{{Physics-informed neural networks modelling for systems with moving immersed boundaries: Application to an unsteady flow past a plunging foil},} J. Fluids Struct. \textbf{125}, 104066 (2024).

\bibitem{Calicchia2023}
M.~A. Calicchia, R.~Mittal, J.-H. Seo, and R.~Ni, \enquote{{Reconstructing the pressure field around swimming fish using a physics-informed neural network},} J. Exp. Biol. \textbf{226}, jeb244983 (2023).

\bibitem{Wang2025}
H.~Wang, F.~Wu, Y.~Liu, X.~He, S.~Feng, and S.~Wang, \enquote{{Machine-learning-based pressure reconstruction with moving boundaries},} J. Fluid Mech. \textbf{1008}, A21 (2025).

\bibitem{Zhu2025}
Y.~Zhu, W.~Chen, J.~Deng, and X.~Bian, \enquote{{Physics-informed neural networks for hidden boundary detection and flow field reconstruction},} arXiv preprint arXiv:2503.24074  (2025).

\bibitem{Zhou2023b}
K.~Zhou and S.~J. Grauer, \enquote{{Flow reconstruction and particle characterization from inertial Lagrangian tracks},} arXiv preprint arXiv:2311.09076  (2023).

\bibitem{Zhou2025}
K.~Zhou, R.~Tang, G.~Ke, and S.~J. Grauer, \enquote{{Neural-implicit particle advection for flow reconstruction from Lagrangian tracks},} in \enquote{{16th International Symposium on Particle Image Velocimetry (ISPIV 2025)},}  (2025), p.~24.

\bibitem{Wang2021}
S.~Wang, Y.~Teng, and P.~Perdikaris, \enquote{{Understanding and mitigating gradient flow pathologies in physics-informed neural networks},} SIAM J. Sci. Comput. \textbf{43}, A3055--A3081 (2021).

\bibitem{Tancik2020}
M.~Tancik, P.~Srinivasan, B.~Mildenhall, S.~Fridovich-Keil, N.~Raghavan, U.~Singhal, R.~Ramamoorthi, J.~Barron, and R.~Ng, \enquote{{Fourier features let networks learn high frequency functions in low dimensional domains},} Adv. Neural Inf. Process. Syst. \textbf{33}, 7537--7547 (2020).

\bibitem{Lee2003}
J.~M. Lee, \emph{{Smooth Manifolds}} (Springer, 2003).

\bibitem{Ma2012}
Y.~Ma and Y.~Fu, \emph{{Manifold Learning Theory and Applications}} (CRC Press, 2012).

\bibitem{Melenk1996}
J.~M. Melenk and I.~Babu{\v{s}}ka, \enquote{{The partition of unity finite element method: basic theory and applications},} Comput. Methods Appl. Mech. Eng. \textbf{139}, 289--314 (1996).

\bibitem{Berkooz1993}
G.~Berkooz, P.~Holmes, and J.~L. Lumley, \enquote{{The proper orthogonal decomposition in the analysis of turbulent flows},} Annu. Rev. Fluid Mech. \textbf{25}, 539--575 (1993).

\bibitem{Taira2017}
K.~Taira, S.~L. Brunton, S.~T. Dawson, C.~W. Rowley, T.~Colonius, B.~J. McKeon, O.~T. Schmidt, S.~Gordeyev, V.~Theofilis, and L.~S. Ukeiley, \enquote{{Modal analysis of fluid flows: An overview},} AIAA J. \textbf{55}, 4013--4041 (2017).

\bibitem{Leake2022}
C.~Leake, H.~Johnson, and D.~Mortari, \emph{{The Theory of Functional Connections: A Functional Interpolation Framework with Applications}} (Lulu.com, 2022).

\bibitem{Berrone2022}
S.~Berrone, C.~Canuto, and M.~Pintore, \enquote{{Variational physics informed neural networks: the role of quadratures and test functions},} J. Sci. Comput. \textbf{92}, 100 (2022).

\bibitem{Mao2023}
Z.~Mao and X.~Meng, \enquote{{Physics-informed neural networks with residual/gradient-based adaptive sampling methods for solving partial differential equations with sharp solutions},} Appl. Math. Mech. \textbf{44}, 1069--1084 (2023).

\bibitem{Wan2024}
X.~Wan, T.~Zhou, and Y.~Zhou, \enquote{{Adaptive importance sampling for Deep Ritz},} Commun. Appl Math. Comput. pp. 1--25 (2024).

\bibitem{Taylor2025}
J.~M. Taylor and D.~Pardo, \enquote{{Stochastic Quadrature Rules for Solving PDEs using Neural Networks},} arXiv preprint arXiv:2504.11976  (2025).

\bibitem{Molnar2025}
J.~P. Molnar and S.~J. Grauer, \enquote{{Algorithm for Time-Resolved Background-Oriented Schlieren Tomography Applied to High-Speed Flows},} in \enquote{{AIAA SciTech 2025 Forum},}  (2025), p. 1060.

\bibitem{Hormann2017}
K.~Hormann and N.~Sukumar, \emph{{Generalized Barycentric Coordinates in Computer Graphics and Computational Mechanics}} (CRC Press, 2017).

\bibitem{Turek2010}
S.~Turek, J.~Hron, M.~Madlik, M.~Razzaq, H.~Wobker, and J.~F. Acker, \enquote{{Numerical simulation and benchmarking of a monolithic multigrid solver for fluid-structure interaction problems with application to hemodynamics},} in \enquote{{Fluid Structure Interaction II: Modelling, Simulation, Optimization},} , H.-J. Bungartz, M.~Mehl, and M.~Sch{\"a}fer, eds. (Springer, 2010), pp. 193--220.

\bibitem{Liu2018}
J.~Liu and A.~L. Marsden, \enquote{{A unified continuum and variational multiscale formulation for fluids, solids, and fluid--structure interaction},} Comput. Methods Appl. Mech. Eng. \textbf{337}, 549--597 (2018).

\bibitem{Updegrove2017}
A.~Updegrove, N.~M. Wilson, J.~Merkow, H.~Lan, A.~L. Marsden, and S.~C. Shadden, \enquote{{SimVascular: an open source pipeline for cardiovascular simulation},} Abbreviation Title Ann. Biomed. Eng. \textbf{45}, 525--541 (2017).

\bibitem{Pan2024}
Y.~Pan and G.~V. Lauder, \enquote{{Combining computational fluid dynamics and experimental data to understand fish schooling behavior},} Integr. Comp. Biol. \textbf{64}, 753--768 (2024).

\bibitem{Menzer2025}
A.~Menzer, Y.~Pan, G.~V. Lauder, and H.~Dong, \enquote{{Fish schools in a vertical diamond formation: Effect of vertical spacing on hydrodynamic interactions},} Phys. Rev. Fluids \textbf{10}, 043104 (2025).

\bibitem{Schanz2016}
D.~Schanz, S.~Gesemann, and A.~Schr{\"o}der, \enquote{{Shake-The-Box: Lagrangian particle tracking at high particle image densities},} Exp. Fluids \textbf{57}, 1--27 (2016).

\bibitem{Mallery2019}
K.~Mallery and J.~Hong, \enquote{{Regularized inverse holographic volume reconstruction for 3D particle tracking},} Opt. Express \textbf{27}, 18069--18084 (2019).

\bibitem{Shao2020}
S.~Shao, K.~Mallery, S.~S. Kumar, and J.~Hong, \enquote{{Machine learning holography for 3D particle field imaging},} Opt. Express \textbf{28}, 2987--2999 (2020).

\bibitem{Moaven2024}
M.~Moaven, A.~Gururaj, V.~Raghav, and B.~Thurow, \enquote{{Improving depth uncertainty in plenoptic camera-based velocimetry},} Exp. Fluids \textbf{65}, 49 (2024).

\bibitem{Molnar2022}
J.~P. Molnar and S.~J. Grauer, \enquote{{Flow field tomography with uncertainty quantification using a Bayesian physics-informed neural network},} Meas. Sci. Technol. \textbf{33}, 065305 (2022).

\end{thebibliography}
\end{document}